\documentclass[12pt]{article}
\usepackage{amsmath}
\usepackage{amsfonts}
\usepackage[mathscr]{euscript}
\usepackage{latexsym}

\renewcommand{\theequation}{\thesection.\arabic{equation}} 
\def\scr{\mathscr}
\def\D{{\scr D}}
\def\Db{\bar\D}

\def\L{{\scr L}}
\def\Lb{\bar\L}
\def\C{{\cal C}}
\newcommand{\N}{{\scr N}}

\def\e{{\rm e}}
\def\d{{\rm d}}

\def\l{\langle}
\def\r{\rangle}
\def\pr{\partial}

\newcommand{\half}{{\textstyle \frac{1}{2}}}
\def\quar{{\textstyle \frac{1}{4}}}
\newcommand{\Leff}{\L_{\rm eff}}
\newcommand{\Leffb}{\Lb_{\rm eff}}

\newcommand{\Geff}{\Gamma_{\rm eff}}
\newcommand{\dS}{\!\!{\rm d}^6z\,}
\newcommand{\dSb}{\!\!{\rm d}^6\bar z\,}
\newcommand{\dV}{\!\!{\rm d}^8z\,}

\newcommand{\al}{\alpha}
\def\da{{\dot\alpha}}
\def\be{\beta}
\def\db{{\dot\beta}}
\def\dg{{\dot\gamma}}
\def\de{\delta}
\def\dd{{\dot\delta}}

\def\is{{^{\!(\sigma)}}}
\def\il{{^{\!(\Lambda)}}}
\def\de{\delta}

\def\ga{\gamma}

\newcommand{\eps}{\varepsilon}
\newcommand{\tfr}[2]{{\textstyle \frac{#1}{#2}}}
\newcommand{\fdq}[2]{\frac{\delta #1}{\delta #2}}

\def\ts{\textstyle}
\newcommand{\dx}{\!\!{\rm d}^4x\,\,}
\def\E{{\cal E}}
\def\A{{\cal A}}
\newcommand{\Liii}{\L_{3}}
\newcommand{\Liiib}{\L_{3}'}

\newcommand{\LRR}{\L_{R\bar R}}
\newcommand{\LW}{\L_{\rm Weyl}}
\newcommand{\Ltop}{\L_{\rm top}}

\newcommand{\IRR}{I_{R \bar R}}
\newcommand{\IW}{I_{\rm Weyl}}
\newcommand{\IWb}{\bar I_{\rm Weyl}}
\newcommand{\Itop}{I_{\rm top}}

\newcommand{\utop}{u_{\rm top}}
\newcommand{\uiii}{u_3}
\newcommand{\uiiib}{u_3'}
\newcommand{\uW}{u_{\rm Weyl}}
\newcommand{\uRR}{u_{R\bar R}}
\newcommand{\rg}{{\rm geom}}

\begin{document}
\thispagestyle{empty}
\begin{flushright}
hep-th/9811209\\ NTZ 34/1998\\ November 1998
\end{flushright}
\begin{center}
{\Large 
 \bf Superconformal Ward Identities for Green Functions}

{\Large \bf  with Multiple Supercurrent Insertions}  

\vspace{1cm}

{\parindent0cm
Johanna Erdmenger\footnote{Supported by Deutsche Forschungsgemeinschaft,
e-mail: Johanna.Erdmenger@itp.uni-leipzig.de} and
Christian Rupp\footnote{Supported by Graduiertenkolleg
  "Quantenfeldtheorie: Mathematische Struktur und physikalische
Anwendungen",
e-mail: Christian.Rupp@itp.uni-leipzig.de}

}

\vspace{1cm}

Institut f{\"u}r Theoretische Physik\\
Universit{\"a}t Leipzig\\
Augustusplatz 10/11\\
D - 04109 Leipzig\\
Germany
\end{center}

\vspace{.5cm}

\centerline{\small \bf Abstract}\vspace*{-2mm} { \small \noindent 
Superconformal Ward identities for N=1 supersymmetric quantum field
theories in four dimensions are convenienty obtained in the superfield
formalism by combining diffeomorphisms and Weyl transformations on curved
superspace. Using this approach we study the superconformal transformation
properties of Green functions with one or more insertions of the supercurrent
to all orders in perturbation theory.
For the case of two insertions we pay particular attention to fixing the
additional counterterms present, as well as to the purely
geometrical anomalies which contribute to the transformation behaviour.
Moreover we show in a scheme-independent way how the quasi-local terms
in the Ward identities are related to similar terms which contribute to the
supercurrent two and three point functions.
 \\ \indent Furthermore we relate our superfield approach to similar studies
which use the component forma\-lism  by discussing the implications of 
our approach for the components 
of the supercurrent and of the super\-gravity prepotentials.
}

\vspace*{10mm}
\begin{tabbing}
\> Curved Superspace
Background, Supercurrent, Anomalies.
\end{tabbing}
\newpage

\section{Introduction}

Superconformal symmetry is an important structure in supersymmetric
theories since the superconformal algebra is a natural extension
of the supersymmetry algebra. In particular, superconformal symmetry
has been used for the construction of correlation functions recently
\cite{f}, \cite{Osborn}. 
In the superfield approach where supersymmetry is manifest,
the central object for investigations of superconformal symmetry
is the supercurrent. This is an axial vector superfield which contains among
its components the $R$ and supersymmetry currents as well as the 
energy-momentum tensor. All currents of the superconformal 
group may be constructed
as space-time moments of the supercurrent. In this respect the supercurrent
plays the same role for supersymmetric theories as the energy-momentum
tensor for non-supersymmetric theories.

To pursue this analogy further it is natural to couple the supercurrent
to the appropriate prepotential of a curved superspace background
within the superfield approach, just as in a non-supersymmetric theory
the energy-momentum tensor
is coupled to the metric $g^{\mu \nu}$
of a curved space background. The combination of superdiffeomorphisms
and super Weyl transformations on curved superspace yields the
superconformal transformations when restricting to flat superspace,
again in close analogy to the non-supersymmetric situation. This
provides a generic method
for deriving superconformal Ward identities, in particular
for Green functions involving the supercurrent. Multiple insertions
of the supercurrent are conveniently generated by varying an appropriate
number of times with respect to the supergravity prepotential. 
 
In quantum field theory, conformal symmetry is expected to be broken 
by hard anomalies in general. These manifest themselves as symmetry breaking
terms in the Ward identities. In a recent paper \cite{ERS1}, it was
shown in a perturbative approach to all orders in $\hbar$ that for
the Wess-Zumino model on curved superspace there are no further dynamical
anomalies than those known from broken scale invariance, characterised
by the $\beta$ and $\gamma$ functions. However there are also purely
geometrical anomalies which are discussed in \cite{ERS2}. These
involve a number of terms, among which are the
supersymmetric Gau\ss-Bonnet and Weyl densities. Since the geometrical
anomalies are of second order in the curvature, they contribute to the
Ward identities for Green functions involving two or more
insertions of the supercurrent.

We use the approach to supergravity presented in \cite{bk}. In addition
to the prepotential coupled to the supercurrent there is also
a chiral compensator which plays an important role for determining
the terms breaking superconformal symmmetry. For a consistent
off-shell formulation we consider also the superconformal transformation
properties of dynamical fields. Similar Ward identities have 
been discussed in \cite{HW1}, though only to first order in the
supergravity fields. Moreover our approach yields the superconformal
transformation properties of the supercurrent in a natural way, thus extending
results  for scalar superfields of \cite{HW2}.
A related though slightly different approach which also considers
Ward identities for double supercurrent insertions may be found in 
\cite{Osborn}.

For a careful treatment of double insertions we consider the
renormalisation of composite operators following the approach of
Zimmermann \cite{quantum}, \cite{zi}. This yields scheme-independent
expressions for Green functions with two or more insertions in a simple way. 
The renormalisation of double insertions
requires additional counterterms involving the external fields,
as discussed in \cite{Shore}. In our case the form of these counterterms is
restricted by the requirement of superdiffeomorphism invariance for the
background supergravity theory.

To illustrate our approach to superconformal
Ward identities, we consider the non-supersymmetric case first in section 2
below and show how classical
conformal Ward identities for the energy-momentum tensor are derived.
In section 3 we introduce diffeomorphisms and Weyl transformations
on curved superspace, emphasising the restrictions which have to be imposed
in order to recover the superconformal transformations when restricting
to flat superspace. In section 4 we derive superconformal Ward identities
and calculate the superconformal transformation properties of the supercurrent
which they imply. In section 5 we discuss the 
anomalies present in the quantised theory. The superconformal transformation
properties of Green functions with one and two insertions ot the supercurrent
are discussed in section 6, paying particular
attention to the contributions arising from purely
geometrical anomalies. In section 7 we discuss the energy-momentum tensor
and further currents
of the supersymmetric theory. By considering the relevant components of
the supercurrent and of the supergravity prepotentials, we show in particular
that the energy-momentum tensor is coupled to the metric as expected. Some
concluding discussions may be found in section 8.

\section{Conformal Transformations  \label{nonsusy}}

\setcounter{equation}{0}

\renewcommand{\mu}{m}
\renewcommand{\nu}{n}

In field theories on flat space, the energy-momentum tensor
$T_{\mu \nu}$ 
is defined as the conserved current associated with translation
invariance. This energy-momentum
tensor may be coupled to the metric of a curved space background according to 
\begin{gather} \label{emt}
T_{\mu \nu} = 2\, \frac{1}{\sqrt{-g}}
\frac{ \delta \Gamma}{\delta g^{\mu \nu}} \, ,
\end{gather}
where $\Gamma$ is the action obtained from the original flat-space
action via the Noether procedure.

On curved space the energy-momentum tensor is
the conserved current associated with diffeomorphism invariance.
Infinitesimal diffeomorphisms are given by the Lie derivative
such that for a coordinate transformation $x_\mu \rightarrow x_\mu - v_\mu$
we have 
\begin{gather}
\delta_v g^{\mu \nu} = \L_v g^{\mu \nu} = v^l \pr_l g^{\mu \nu}
- \pr_l v^\mu g^{\nu l} - \pr_l v^\nu g^{\mu l}
= -\nabla^\mu v^\nu - \nabla^\nu v^\mu \, 
\end{gather} 
for the infinitesimal transformation of the metric.

For 
Weyl transformations or local rescalings, the 
infinitesimal transformation of the metric is given by 
\begin{gather}
\delta_\sigma g^{\mu \nu} = - 2 \sigma g^{\mu \nu}
\end{gather}
with $\sigma(x)$ an arbitrary function. Requiring the metric
to be invariant under the combination of diffeomorphisms
and Weyl transformations we have
\begin{gather} \label{Liemetric}
0 = - (\delta_v + \delta_\sigma) g^{\mu \nu} = 
\nabla^\mu v^\nu + \nabla^\nu v^\mu + 2 \sigma g^{\mu \nu} \, .
\end{gather}
This requires that $\sigma$ is restricted to
$\sigma =\sigma_{(v)}$ in four dimensions, with
\begin{gather} \label{sv}
\sigma_{(v)} = - {\ts \frac{1}{4}} \nabla \cdot v \, .
\end{gather}
On flat space 
(\ref{Liemetric}) becomes
\begin{gather} \label{confeqn}
\pr_\mu v_\nu + \pr_\nu v_\mu = -2 \sigma_{(v)} \eta_{\mu \nu} \, ,
\end{gather}
which is the well-known conformal Killing equation defining conformal
transformations on flat space.

The diffeomorphism transformation
properties of the
field theory given by $\Gamma$ 
may be studied by means of an appropriate Ward identity.
For example for an  action $\Gamma$ involving 
a dynamical scalar field $\varphi$ besides the metric,  we have
a diffeomorphism Ward identity of the form
\begin{gather}  
 \; \delta_v \Gamma [g, \varphi] 
= \int \! \d^{\, 4} x \, \left( {\L}_v g^{\mu \nu}  
\frac{\delta \Gamma }
{\delta g^{\mu \nu}} + {\L}_v \varphi  \frac{\delta \Gamma}
{\delta \varphi}  \right) 
\equiv \int \! \d^{\,4} x \, \sqrt{-g} \, v^\mu w^{(v)}{}_\mu (g, \varphi)
\Gamma \,    \, ,
\label{diffeo} \end{gather}
where we have used (\ref{Liemetric}) and (\ref{emt}) as well as
$g^{\mu \nu} = g^{(\mu \nu)}$, $\L_v \varphi = v^\mu \pr_\mu \varphi$, and
\begin{gather}
w^{(v)}{}_\mu(g,\varphi) \equiv w^{(v)}{}_\mu(g) + w^{(v)}{}_\mu(\varphi) \, , 
\\ {\rm where} \;
w^{(v)}{}_\mu(g) = 2 \nabla^\nu \left( (-g)^{- \frac{1}{2}}
\frac{\delta}{\delta g^{\mu \nu}} \right)
\, , \; \; \; w^{(v)}{}_\mu(\varphi) = (-g)^{- \frac{1}{2}} \,
 \pr_\mu \varphi \frac{\delta}{\delta \varphi} \, ,  \nonumber
\end{gather}
is a local Ward operator.  
If $\Gamma$ is diffeomorphism invariant, $\delta_v \Gamma [g,\varphi]=0$, 
we have using  (\ref{emt})
\begin{gather} \label{wv}
  \, w^{(v)}{}_\mu(\varphi) \Gamma =
- \nabla^\nu T_{\mu \nu} \, 
\end{gather}
since $v^\mu(x)$ is arbitrary.
Therefore diffeomorphism invariance implies conservation of the energy-momentum
tensor since $w^{(v)}{}_\mu(\varphi) \Gamma$ 
vanishes if the equations of motion
are imposed.

The Weyl transformations may be discussed in a similar fashion.
For the  Weyl transformation of the  action $\Gamma [g,\varphi]$ we have
\begin{gather} 
 \delta_\sigma \Gamma [g, \varphi] 
= \int \! \d^{\, 4} x \,  \left( \delta_\sigma g^{\mu \nu}   
\frac{\delta \Gamma}
{\delta g^{\mu \nu}} + \delta_\sigma \varphi   \frac{\delta \Gamma}
{\delta \varphi} \right)
 \equiv \int \! \d^{\, 4} x\, \sigma w^{(\sigma)} \Gamma \, ,
\label{Weylt} \end{gather}
where
\begin{gather}
w^{(\sigma)}(g,\varphi) = - 2 g^{\mu \nu} \frac{\delta}{\delta g^{\mu \nu}}
-  d_\varphi \varphi {\frac{\delta}{\delta \varphi}}
\, ,
\end{gather}
and $d_\varphi$ is the scale dimension of $\varphi$.
We allow for the action $W$  not to be Weyl invariant in general. 
The breakdown of Weyl symmetry is then expressed by the local
Ward identity
\begin{gather} \label{ws}
(-g)^{- \frac{1}{2}} w^{(\sigma)}(g, \varphi) \Gamma = - \, T \, ,
\end{gather}
with $T$ a scalar density.
Using (\ref{emt}) the imposition of the equation of motion
yields
\begin{gather}
T^\mu{}_\mu = T
\end{gather} 
such that Weyl invariance, $T=0$, is equivalent to the tracelessness
of the energy-momentum tensor.

For Weyl transformations for which $\sigma$ is restricted by (\ref{sv}),
the transformations
under diffeomorphisms and Weyl transformations may be combined into
one single equation by adding (\ref{wv}) and (\ref{ws}). Since
now the arbitrary function $v$ is common to both transformations we have  
\begin{align}
\delta_v  \Gamma [g,\varphi] + \delta_\sigma  \Gamma [g,\varphi] &= 
\int \! \d^{\,4} x \, \sqrt{-g} v^\mu \, w_\mu (g, \varphi) \Gamma  \, , \\
w_\mu (g, \varphi) \Gamma &= - {\ts \frac{1}{4}} \nabla_\mu T \, , \label{cs}
\end{align}
\begin{align}
\;\; {\rm where} \;\;\;\;
w_\mu(g,\varphi) &= 2\nabla^\nu ( (-g)^{- \frac{1}{2}}
\frac{\delta}{\delta g^{\mu \nu}} ) -
{\ts \frac{1}{2}} \nabla_\mu ( (-g)^{- \frac{1}{2}} g^{k l}   \frac{\delta}
{\delta g^{k l}} ) \nonumber\\ & \;\;\;\; +  (-g)^{- \frac{1}{2}} \,
\pr_\mu \varphi \frac{\delta}{\delta \varphi} - \frac{d_\varphi}{4} \pr_\mu
( (-g)^{- \frac{1}{2}} \, \varphi \frac{\delta}{\delta \varphi}) \,. 
\end{align}
(\ref{cs}) is referred to as the conformal Ward identity.
When restricting to flat space, (\ref{cs}) reduces to
\begin{gather} \label{nsc}  w_\nu(\varphi) \Gamma =
- \pr^\mu ( T_{\mu \nu} - {\ts \frac{1}{4}} \eta_{\mu \nu}
 T^l{}_l )  
  - {\ts \frac{1}{4}} \pr_\nu T \, ,
\end{gather}
where the Ward operator
\begin{gather}
w_\mu(\varphi) = 
\pr_\mu \varphi \frac{\delta}{\delta \varphi} - \frac{d_\varphi}{4} \pr_\mu
( \varphi \frac{\delta}{\delta \varphi})
\end{gather} 
generates the conformal transformations of the flat space theory in a
consistent off-shell formalism. 
The r.h.s.~of the flat space conformal Ward
identity (\ref{nsc}) consists of the divergence of the improved traceless
current $T'{}_{\mu \nu} \equiv T_{\mu \nu} - {\ts \frac{1}{4}} \eta_{\mu \nu}
 T_l{}^l$ and of the trace term $T$ which breaks the conformal
symmetry.

In quantum field theories conformal symmetry is expected to be broken
by anomalies. It is desirable to impose diffeomorphism invariance
such that the energy-momentum tensor remains conserved. 
Then the breakdown of conformal
symmetry is entirely determined by the breakdown of Weyl symmetry.

\newpage

\section{Superconformal Transformations}

\setcounter{equation}{0}

Let us now consider the supersymmetric case\footnote{In our conventions 
the metric has diagonal elements $(+1,-1,-1,-1)$.The covariant derivatives
satisfy $\{ D_\al, \bar D_\da\} = 2 i \sigma^a_{\al \da} \pr_a$.}. 
In analogy to the 
previous discussion, superconformal transformations are given by
the flat space restriction of the combination of superdiffeomorphisms
and super Weyl transformations on curved superspace.
Within the superfield formalism we consider the class of supergravity
theories which are characterised by two prepotentials, the vector superfield
$H$ and the chiral compensator $\phi$ as well as its chiral conjugate
$\bar \phi$.  These fields, which are all dimensionless, play the same role
as the metric $g^{m n}$ in the non-supersymmetric case.
We note that there are two  geometrical fields in this
case, whereas there is only one in the non-supersymmetric situation.
The discussion below demonstrates which properties of  $g^{m n}$
are attributed to the two independent fields $H$ and $\phi$ respectively.

The real axial vector superfield $H$ may be expanded in the basis
\begin{gather} \label{H}
H = H^a \pr_a + H^\alpha D_\alpha + \bar H_\da \bar D^\da 
+ H^{\al \beta} M_{\alpha \beta} + \bar H^{\da \db} M_{\da \db} \, ,
\end{gather}
where
\begin{gather} \label{basis}
D_A \equiv (\pr_a, D_\al, \bar D^\da )
\end{gather}
are the well known partial and flat space supersymmetry derivatives
respectively, which span the tangent space. $M_{\al \beta}$ and $M_{\da \db}$
are the generators of infinitesimal local Lorentz transformations, which
act on spinor indices only. 
The chiral compensator is a scalar which satisfies
the chirality constraint
\begin{gather}
\bar D_\da \phi = 0 \, .
\end{gather}
The field $H$ is at the heart of the construction of supercovariant
derivatives, which are the covariant derivatives associated
to the group of superdiffeomorphisms on curved superspace. These derivatives
are denoted by
\begin{gather}
\D_A \equiv ( \D_a, \D_\al, \bar \D^{\da} ) \, ,
\end{gather}
which reduces to (\ref{basis}) in the flat superspace limit given by
$H=0$, $\phi=1$. The details of the construction of the supercovariant 
derivatives may be found in chapter 5 of the book \cite{bk} (see also paper I).
It should be noted that throughout this paper we use the curved space
chiral representation
which is analogous to the flat space chiral representation in which all fields
$\tilde \Phi$ are replaced by 
$\Phi = \e^{i\theta\sigma^a\bar\theta\partial_a} \tilde\Phi$.
Complex conjugation of $\Phi$ in the chiral representation yields $\bar \Phi$
in the antichiral representation. On curved space, the chiral 
representation expression for the conjugate of $\Phi$ 
is given by  ${\rm e}^{2iH} \bar \Phi$.

The superdiffeomorphisms are given by a complex superfield
$\Lambda$ and its conjugate $\bar \Lambda$, which may be expanded in the
basis of (\ref{H}),
\begin{align}
\Lambda &= \Lambda^a \pr_a + \Lambda^\alpha D_\alpha + \Lambda_\da \bar D^\da 
+ \Lambda^{\al \beta} M_{\alpha \beta} + \Lambda^{\da \db} M_{\da \db} \, ,
\nonumber\\
\bar \Lambda &= \bar \Lambda^a \pr_a + \bar \Lambda^\alpha D_\alpha + \bar 
\Lambda_\da \bar D^\da 
+ \bar \Lambda^{\al \beta} M_{\alpha \beta} + 
\bar \Lambda^{\da \db} M_{\da \db} \, , \label{diffeom}
\end{align}
where $\Lambda$ is in the chiral and $\bar \Lambda$ in the 
antichiral representation. The diffeomorphisms $\Lambda$ and $\bar \Lambda$
are restricted by requiring the covariant transformation laws
\begin{equation}
\D_A  \longrightarrow  \e^\Lambda  \D_A \e^{-\Lambda} \, , \;\;\;\;
\D_A  \longrightarrow  \e^{- \bar\Lambda} \D_A \e^{\bar \Lambda}
\end{equation}
for the supercovariant derivatives.
These requirements impose the constraints
\begin{gather} \label{Lconstraint}
\bar D_{\dot\beta}\Lambda^{\alpha\dot\alpha} = 4i \Lambda^\alpha 
\delta_{\dot\beta}^{\dot\alpha}, \;\;\;\;
\bar D_{\db} \Lambda^\alpha =0, \;\;\;\; 
\Lambda_{\dot\alpha\dot\beta} = -\bar D_{(\dot\alpha} \Lambda_{\dot\beta)} 
\, ,
\end{gather}
and similar relations for the components of $\bar \Lambda$ which may
be obtained by complex conjugation.
The diffeomorphism transformation properties of the prepotential $H$
are given by
\begin{gather} \label{Htrans}
\e^{2iH}  \longrightarrow  \e^\Lambda \e^{2iH} \e^{-\bar \Lambda} \, ,
\end{gather}
which are analogous to the gauge transformations  of a non-abelian gauge field.
By imposing the additional constraints
\begin{equation} \label{H2}
\Lambda_{\dot\alpha}= \e^{2iH} \bar \Lambda_{\dot\alpha} \, , 
\;\;\;\;
\Lambda_{\alpha\beta} = \e^{2iH} \bar \Lambda_{\alpha\beta}\, , 
\end{equation}
$H$ may be reduced to the simpler form
\begin{equation} \label{Hsimple}
H = H^{\alpha\dot\alpha} D_{\alpha\dot\alpha} \, ,
\end{equation}
where all other components of $H$ vanish.
The diffeomorphism transformation properties of the chiral
compensator are given by
\begin{gather} \label{phi2}
\phi^3 \rightarrow  \phi^3 \, \e^{\overleftarrow{\Lambda_c}} \, , \qquad
\Lambda_c = \Lambda^a D_a + \Lambda^\al D_\al 
\end{gather}
in the chiral representation, and
\begin{gather}
\bar \phi^3 \rightarrow  \bar \phi^3 \, \e^{\overleftarrow{\bar\Lambda_c}}
 \, , \qquad
\bar \Lambda_c = \bar \Lambda^a D_a + \bar \Lambda_\da \bar D^\da
\end{gather}
in the antichiral representation.
The constraints (\ref{Lconstraint}) may be solved
in terms of a complex superfield $\Omega^\alpha$,
\begin{equation}
\Lambda^{\alpha\dot\alpha} = i \bar D^{\dot\alpha} \Omega^\alpha\, ,\;\;\;\;
\Lambda^\alpha = \tfr{1}{4} \bar D^2 \Omega^\alpha \, .
\label{omegadef}
\end{equation}
Equation \eqref{omegadef} determines $\Omega^\alpha$ uniquely up to 
a chiral field. Correspondingly for $\bar \Lambda$ we have
\begin{equation} 
\bar \Lambda^{\alpha\dot\alpha} = i  D^{\alpha} \bar \Omega^\da \, ,\;\;\;\;
\bar \Lambda^\da = \tfr{1}{4}  D^2 \bar \Omega^\da \, .
\label{baromegadef}
\end{equation}
(\ref{H2}) and
its conjugate ensure that $\Lambda_\da$ is a function of $\bar \Omega_\da$
and $\bar \Lambda^\al$ is a function of $\Omega^\al$.

The super Weyl transformations are given by
superfields $\sigma$, $\bar \sigma$ which satisfy $\bar D_\da \sigma = 0$,
$ D_\al \bar \sigma = 0$. For the Weyl transformation properties
of the prepotentials we have
\begin{gather} \label{W2}
H \longrightarrow H , \qquad
\phi \longrightarrow \e^\sigma \phi , \qquad 
\bar \phi \longrightarrow \e^{\bar \sigma} \bar \phi.
\end{gather}
It is crucial to note that $H$ is a Weyl invariant.

We now perform the same analysis as discussed above for the non-supersymmetric
case and combine the infinitesimal diffeomorphisms and Weyl transformations
of the prepotentials. Restricting to flat superspace we thus obtain the
superconformal transformations.
The infinitesimal transformations are obtained by expanding (\ref{H2}), 
(\ref{phi2}) and (\ref{W2}) for small  $\Lambda, \bar \Lambda, \sigma, \bar 
\sigma$, which yields
\begin{gather} \label{conftrans}
\delta H = \delta_{\Lambda, \bar\Lambda}
 H \, , \qquad
\delta \phi = \delta_\Lambda \phi + \delta_\sigma \phi \, , \qquad
\delta \bar \phi = \delta_{\bar \Lambda} \bar \phi + \delta_{\bar \sigma}
 \bar \phi \, ,
\end{gather}
where
\begin{gather}
2i \delta_{\Lambda,
\bar \Lambda} H = \Lambda-\bar\Lambda - {\ts\half} [2iH, 
\Lambda+\bar \Lambda]
+{\ts\frac{1}{12}} [2iH, \left[2iH, \Lambda-\bar \Lambda] \right] + O(H^3) 
 \, , \nonumber\\
\label{delta_Lambda_H}
\delta_\Lambda \phi = \Lambda \phi + {\textstyle \frac{1}{3}} \left(
D^{\alpha \da} \Lambda_{\alpha \da} + D^\alpha \Lambda_\alpha \right) \phi \, ,
 \qquad 
\delta_{\bar \Lambda} \bar \phi = \bar \Lambda \bar \phi 
+ {\textstyle \frac{1}{3}} \left(
D^{\alpha \da} \bar \Lambda_{\alpha \da} + \bar D_\da \bar \Lambda^\da 
\right) \bar \phi
 \, , \qquad 
 \\
\delta_\sigma \phi =  \sigma \phi \, , \qquad
\delta_{\bar \sigma} \bar \phi = \bar \sigma \bar \phi 
 \label{delta_sigma_AJH} \, .
\end{gather}
The three independent conditions
\begin{subequations}
\begin{gather}
0 = \delta H \, , \\
0 = \delta \phi \, , \qquad 
0 = \delta \bar\phi \, ,
\end{gather}
\end{subequations}
for the variations of (\ref{conftrans}), 
which extend (\ref{Liemetric}) to the supersymmetric case,
require that
to lowest order in the prepotentials
\begin{subequations}
\label{Lsflach}
\begin{gather} \label{Lflach}
\Lambda =\bar\Lambda \, , \\ \label{sflach}
\sigma = -{\ts\frac{1}{3}}  
\left(D^{\al\da} \Lambda_{\al\da}+ D^\al \Lambda_\al
\right)  \, , \qquad 
\bar \sigma = -{\ts\frac{1}{3}}  
\left(D^{\al\da} \bar \Lambda_{\al\da}+ \bar D_\da \bar \Lambda^\da
\right)   \, .
\end{gather}
\end{subequations}
These conditions are the superspace analogue of (\ref{sv}). 
In terms of $\Omega, \bar \Omega$,  these three conditions are given by
\begin{subequations}
\label{oLsflach}
\begin{gather} \label{oLflach}
\bar D^\da \Omega^\al = D^\al \bar \Omega^\da \, , \\ \label{osflach}
\sigma = - {\ts \frac{1}{12}} \bar{D}^2 D^\alpha \Omega_\alpha \, , \qquad
\bar \sigma = - {\ts \frac{1}{12}} {D}^2 \bar D_\da \bar \Omega^\da \, .
\end{gather}
\end{subequations}
(\ref{oLflach}) is on its own a necessary and sufficient condition for
$\Omega$ to contain the superconformal coordinate transformations. 
The most general solution for $\Omega$ satisfying 
(\ref{oLflach})  is given by
\begin{align}
\Omega^\al(z) &= -\tfr{i}{2} \sigma_a{}^{\al\da} \bar\theta_\da \,
v^a(x) + \tfr{1}{8} \left( \partial^a v_a - 8i\,r \right)
\nonumber\\
&\quad - \bar\theta^2 \zeta^\al(x) + 2 \theta^\al \bar\theta_\da \bar
\zeta^\da(x) \label{omegasol}
\end{align}
with
\begin{align}
v^a(x) &= t^a -  
(\omega^{ab}-\omega^{ba})\,x_b + d\,x^a  - k^a \,x^2 + 2 k^b \, x_b
x^a \,,\nonumber\\
\zeta^\al(x) &= q^\al + \sigma^{a\al\da} \bar s_\da x_a \,. \nonumber
\end{align}
$v^a(x)$ satisfies the conformal Killing equation (\ref{confeqn}).
The term with parameter $r$ in (\ref{omegasol}) corresponds to the $R$
transformations, the terms in $v^a$ correspond to translations ($t^a$),
Lorentz transformations ($\omega^{ab}$), dilatations ($d$) and special
conformal  transformations ($k^a$) respectively.
$q^\al$ and $\bar s_\da$ in  $\zeta^\al$ are the parameters of
supersymmetry and special supersymmetry transformations.
For $\Omega$ given by (\ref{omegasol}) the Weyl
transformation parameter $\sigma$ in (\ref{osflach}) 
is given by
\begin{gather} \label{sigmacoeff}
\sigma =  {\ts \frac{2}{3}} i r - d + 4i s^\al \theta_\al - 2 k^a x_a
+ 2i \theta \sigma^a \bar \theta k_a \, , 
\end{gather}
such that $\sigma$ is non-zero only for the $R$ transformations, dilatations,
special supersymmetry and special conformal transformations.

The imposition of the conditions (\ref{osflach}) allows to combine
the infinitesimal diffeomorphism and Weyl transformations of the prepotentials
and also of possible dynamical fields
into a single transformation even on curved space, independently of
whether (\ref{oLflach}) is imposed or not. For vector superfields like
$H$ the combined transformation depends on both $\Omega, \bar \Omega$
whereas for chiral fields like $\phi$ the combined transformation
depends only on $\Omega$. For antichiral fields like $\bar \phi$ it depends
only on $\bar \Omega$.

\section{Supercurrent and Ward identities}

\setcounter{equation}{0}

The supercurrent $V_a$ is the real axial
vector superfield which contains the R and supersymmetry currents
as well as the energy-momentum tensor among its components. 
It is the basic ingredient for constructing all currents of
the superconformal group. 
On curved superspace it is natural to couple the supercurrent
to the supergravity field $H_{\al \da}$ as given by (\ref{Hsimple}),
\begin{gather} \label{sch}
V_{\al \da} = 8  \, \frac{ \delta \Gamma
}{\delta H^{\al \da}} \, ,
\end{gather}
in the same way as the energy-momentum tensor is coupled to
the metric in (\ref{emt}). A difference to the non-supersymmetric case
is that here there is no need for a density factor involving a determinant.
This is due to the fact that we perform a so-called `background-quantum
splitting' \cite{bk}, \cite{ggrs}. The background around which we expand
in terms of $H_{\al \da}$ 
is flat superspace for which the determinant is equal to one. 
It should be noted that the expression 'background-quantum splitting'
is misleading in our context since $H_{\al \da}$ is a classical field.

Just as in section \ref{nonsusy}  we may consider diffeomorphism 
and Weyl symmetry Ward identities. 
For later convenience it is useful to define a chiral dimensionless
field $J$ by
\begin{align}
\phi &\equiv \e^J = 1 + J + \half J^2 + \dots \, ,
\end{align}
such that flat superspace is characterised by vanishing external
fields $H=0, J=0$ rather than by $H=0, \phi=1$. From its definition
and the transformation properties of $\phi$ the infinitesimal
transformations of $J$ are given by
\begin{gather}
\delta J = \Lambda J + {\ts \frac{1}{3}} (D^{\alpha \da} \Lambda_{\al
\da} + D^\al \Lambda_\al ) + \sigma \, .
\end{gather}

For the Ward identities we consider as an example a Wess-Zumino type action
\newline $\Gamma [H,\phi,\bar \phi, A, \bar A]$ on curved superspace, with
chiral matter fields $A$. In addition to the transformation properties
(\ref{conftrans}) for the prepotentials we have
\begin{gather}
\delta_\Lambda A = \Lambda A \, , \;\; \delta_\sigma A = - \sigma A \, , \qquad
\delta_{\bar \Lambda} \bar A = \bar\Lambda \bar A \, , \;\; 
\delta_{\bar \sigma} \bar A = - \bar \sigma \bar A \, 
\end{gather}
for the diffeomorphism and Weyl transformation properties of the 
dynamical fields $A$, $ \bar A$.
For the diffeomorphisms local Ward operators $w^{(\Lambda)}_\al(H)$, $\bar
w^{(\Lambda)\da}(H)$ are defined by
\begin{align}
\int \! \d^{\, 8 } z \,  \delta_{\Lambda, \bar \Lambda} H^{\al \da}
\frac{\delta \Gamma} {\delta H^{\al \da}} & = 
\int\dV \delta_\Omega H^{\al\da} \fdq{\Gamma}{H^{\al\da}} + \int\dV \delta_{\bar
    \Omega} H^{\al\da} \fdq{\Gamma}{H^{\al\da}}  \nonumber \\
&= \int \dV
\Omega^\al w^{(\Lambda)}{}_\al (H) \Gamma + \int \dV
\bar \Omega_\da \bar w^{(\Lambda)}{}^\da (H) \Gamma \,  
\label{wHdef}
\end{align}
in an expansion around flat space,
where (\ref{H2}), (\ref{omegadef}) and (\ref{baromegadef}) have been 
used to express $\Lambda$, $\bar \Lambda$  in terms of $\Omega$, $\bar \Omega$.
The explicit expression for $w^{(\Lambda)}{}_\al (H)$ and all other
local Ward operators is given in appendix \ref{wardoperators}. In an expansion 
around flat space, the operator 
$w^{(\Lambda)}{}_\al (H)$, which incorporates the
transformation law (\ref{Htrans}), is given by a power series
in $H^{\al \da}$.
For $J$,$\bar J$ the local diffeomorphism Ward operators are defined by
\begin{gather}
\int \! \d^8z\,\, \delta_\Lambda J \frac{\delta \Gamma}
{\delta J} = \int \! \d^8z\,\, 
\Omega^\al \, w_\al\il(J)  \Gamma
   \, , \qquad 
\int \! \d^8z\,\, \delta_{\bar \Lambda} \bar J \frac{\delta \Gamma}
{\delta \bar J} = \int \! \d^8z\,\, 
\bar \Omega_\da \, w^\da\il(\bar J)  \Gamma \, ,
\end{gather}
and for the dynamical fields by
\begin{gather}
\int \! \d^8z\,\, \delta_\Lambda A \frac{\delta \Gamma}
{\delta A} = \int \! \d^8z\,\, 
\Omega^\al \, w_\al\il(A)  \Gamma
   \, , \qquad 
\int \! \d^8z\,\, \delta_{\bar \Lambda} \bar A \frac{\delta \Gamma}
{\delta \bar A} = \int \! \d^8z\,\, 
\bar \Omega_\da \, w^\da\il(\bar A)  \Gamma \, .
\end{gather}
For a diffeomorphism invariant action $\Gamma$ we have
\begin{gather} \label{difW}
w_\al\il(H,J,A) \Gamma \equiv 
\left( w_\al\il (H) + w_\al\il(J) + w_\al\il(A) \right) \Gamma = 0 \, ,
\\ 
\bar w_\da\il(H,\bar J, \bar A) \Gamma \equiv 
\left( \bar w_\da\il (H) + \bar w_\da\il(\bar J) + w_\da\il(\bar A) \right) 
\Gamma = 0 \, ,
\end{gather}
since $\Omega$, $\bar \Omega$ are independent and arbitrary.

Similarly for the Weyl transformations we have
\begin{gather} \label{wis}
\int \! \d^6 z \,\, \left( \delta_\sigma A \frac{\delta \Gamma }{\delta A}
\,+ \, \delta_\sigma \phi \frac{\delta \Gamma }{\delta \phi} \right) =
\int \! \d^6 z \,\, \sigma \,  w\is(A,J) \Gamma \, , \\
w\is(A,J) = \frac{\delta}{\delta J} - A \frac{\delta}{\delta A} \, .
\nonumber
\end{gather}
from which the expression for $\bar w\is(\bar A, \bar J)$ is obtained
by complex conjugation. The action $\Gamma$ is not Weyl invariant in general.
Terms breaking Weyl symmetry are denoted by $S$, $\bar S$,
\begin{gather} \label{AJswi}
w\is(A,J) \Gamma = - {\ts \frac{3}{2}} S \, ,  \qquad  \,
\bar w\is(\bar A,\bar J) \Gamma = - {\ts \frac{3}{2}} \bar S   \, .
\end{gather}  
For Weyl transformations restricted by (\ref{osflach}) we have
\begin{align}
\int \! \d^6z \, \sigma \, w\is(A,J)\, 
 \Gamma &= \, \int \! \d^8z \, 
\Omega^\al w_\al\is(A,J) \, \Gamma \\ 
{\rm with} \qquad
w_\al\is(A,J) &= \,  {\ts \frac{1}{12}} D_\al w\is(A,J) \, , \label{walsi}
\end{align}
which yields 
\begin{align} \label{swi}
w_\al\is(A,J) \Gamma &= - {\ts \frac{1}{8}} D_\al S \, . 
\end{align}
For the antichiral transformation with parameter $\bar \sigma$, similar
expressions are obtained by complex conjugation again.
The Weyl identity given by (\ref{walsi}) may be combined with
the diffeomorphism identity  (\ref{difW}),
\begin{gather} w_\al(A,H,J)  \Gamma \equiv
\left( w_\al\il(A,H,J) + w_\al\is(A,J) \right) 
\Gamma = - {\ts \frac{1}{8}} D_\al S  \label{combiW}
\end{gather}
with $w_\al(A,H,J)$ the Ward operator for the combined chiral diffeomorphism
and Weyl transformations of $H$, $J$ and $A$.
When restricting to flat space, $H=0$, $J=0$ (\ref{combiW}) yields 
\begin{align} \label{susych}
-16 w_\al(A)  \Gamma &= \bar D^\da V_{\al \da} + 2 D_\al S \,
\end{align}
with $w_\al(A)$ the combined Ward operator for  diffeomorphism and Weyl
transformations of the dynamical field $A$. This equation may be combined
with its conjugate to yield
\begin{align} \label{sctt}
8 i (D^\al w_\al - \bar D_\da \bar w^\da)\Gamma   &= \pr^a V_a
- i ( D^2 S - \bar D^2 \bar S) \, .
\end{align}
By combining (\ref{susych}) with its conjugate we have tacitly assumed
the validity of 
the condition for conformal transformations, $\bar D^\da \Omega^\al
= D^\al \bar \Omega^\da$, as given by (\ref{oLflach}). 
Therefore
(\ref{sctt})
describes the superconformal transformation properties of the flat
space theory and thus is the supersymmetric equivalent of (\ref{nsc}).

With these results we may obtain the transformation properties of the
supercurrent. For independent $\Omega$, $\bar \Omega$ we obtain 
by varying (\ref{combiW}) and its conjugate separately with respect
to $H^{\be \db}(z')$, using (\ref{sch}) 
and the expressions of appendix \ref{wardoperators},  
and subsequently restricting to flat space
\begin{subequations}
\begin{align}
w_\al(A)(z) V_{\be \db}(z') = &  - \tfr{1}{16} \bar D^\da
G_{\al\da\be\db}(z,z') \nonumber\\ & 
- \tfr{1}{4} \Big( \bar D^\da ( 
\{ D_\al , \bar D_\da\} \delta^8(z-z') V_{\be \db} ) + \{ D_\be ,\bar D_\db\}
\bar D^\da ( \delta^8(z-z') V_{\al \da} ) \nonumber\\ 
& \hspace{3cm} + \bar D^2 (D_\al \delta^8 (z-z')
V_{\be \db}) \Big)  
\nonumber\\ & 
-  D_\al \frac{\delta S(z)}{\delta H^{\be \db} (z')} \Big|_{H=J=0} \, ,
\label{wic} \\ 
& \nonumber\\
\bar w_\da(\bar A)(z) V_{\be \db}(z') 
= &  - \tfr{1}{16}  D^\al  G_{\al\da\be\db}(z,z') \nonumber\\ & 
+\tfr{1}{4} \Big(  D^\al ( 
\{ D_\al , \bar D_\da\} \delta^8(z-z') V_{\be \db} ) + \{ D_\be ,\bar D_\db\}
 D^\al ( \delta^8(z-z') V_{\al \da} ) \nonumber\\ 
& \hspace{3cm}-  D^2 (\bar D_\da \delta^8 (z-z')
V_{\be \db}) \Big)  \nonumber\\ & 
-  \bar D_\da \frac{\delta \bar S(z)}{\delta H^{\be \db} (z')}
\Big|_{H=J=0} \, , \label{wiac} 
\end{align}
\end{subequations}
where we have defined 
\begin{gather}
G_{\al\da\be\db}(z,z') \equiv
64 \frac{\delta^2 \Gamma}{\delta H^{\al \da} (z) \delta H^{\be \db}(z')}
\Big|_{H=0, J=0} \, .  
\end{gather}

Furthermore our construction enables us to derive the superconformal
transformation properties of the supercurrent for each of the different
superconformal coordinate transformations explicitly. For this we
consider the local curved space 
Ward identity (\ref{combiW}) in its integrated form, 
\begin{gather} \label{integW}
W(A,H,J) \Gamma
 = - \tfr{1}{8} \, \int \d^{\, 8} z \, \left( \Omega^\al D_\al S
+ \bar \Omega_\da \bar D^\da \bar S \right) \, , \\
W(A,H,J) = \int\dV \left( \Omega^{\al} w_\al(A,H,J) \,+\, 
\bar \Omega_\da \bar w^\da (\bar A,H,\bar J) \right)\,. \label{Wop}
\end{gather}
By varying this equation with respect to $H^{\al \da} (z)$, using
$\bar D^\da \Omega^\al = D^\al \bar \Omega^\da$ such as to obtain
the superconformal transformations, and restricting to flat space
we obtain
\begin{align} 
W(A) V_{\al\da} 
& = \, \quar \left( \left\{ D_\beta, \bar D_\db \right\}
\big( \bar D^\db \Omega^\beta V_{\al \da} \big) + \big(\left\{ D_\al,
\bar D_\da \right\} \bar D^\db \Omega^\beta \big) V_{\beta \db} + D^
\beta \left( \bar D^2 \Omega_\beta V_{\al \da} \right) \right)
\nonumber \\
& \quad \, - 12 \int\dS \sigma \fdq{S}{H^{\al\da}}
\,+\,c.c. \, . \label{Vtransclass}
\end{align}
In the superconformal case where the $S$ terms vanish we have
\begin{align}
\delta V_{\al\da} &=
 \quar \left( \left\{ D_\beta, \bar D_\db \right\} 
 \big( \bar D^\db \Omega^\beta V_{\al \da} \big) +
\big(\left\{ D_\al, \bar D_\da \right\} \bar D^\db \Omega^\beta 
\big) V_{\beta \db} + D^ \beta \left( \bar D^2 \Omega_\beta V_{\al \da}
\right) \right) +c.c. \nonumber 
\\
&= \left( t^a \, \delta^P_a + q^\al \, \delta^Q_\al + \bar q_\da \,
 \delta^{\bar Q\da} + \omega^{ab} \, \delta^M_{ab} + r \, \delta^R + d
 \, \delta^D + s^\al \, \delta^S_\al + \bar s_\da \, \delta^{\bar
 S\da} + k^a \, \delta^K_a \right) \, V_{\al\da} \, , 
\label{deltaclassV}
\end{align}
where we have inserted $\Omega$ as given by
(\ref{omegasol}) to obtain the transformation
properties for the different superconformal coordinate transformations
respectively, which yields
\begin{align}
\delta^P_a \, V_{\al\da} &= \partial_a \,V_{\al\da}\,, \nonumber\\[0.3cm]
\delta^Q_\ga \, V_{\al\da} &= \left(\partial_\ga + i \sigma^a_{\ga\dg}
  \bar\theta^\dg \partial_a \right) \, V_{\al\da}\,, \nonumber\\[0.3cm]
\delta^{\bar Q \dg} \,V_{\al\da}&= \left( -\bar\partial^\dg -i \theta^\ga
  \sigma^a{}_\ga{}^\dg \partial_a \right) \, V_{\al\da}\,, \nonumber\\[0.3cm]
\delta^M_{ab} \, V_{\al\da} &= \left( x_a\partial_b - x_b\partial_a
  -\tfr{i}{2} \theta^\be (\sigma_{ab})_\be{}^\ga \partial_\ga +
  \tfr{i}{2} \bar\theta_\db (\bar\sigma_{ab})^\db{}_\dg \bar\partial^\dg\right) \,
V_{\al\da} \nonumber\\
  &\quad -\tfr{i}{2}
  \left( (\sigma_{ab})_\al{}^\be \delta_\da{}^\db -
     \delta_\al{}^\be (\bar\sigma_{ab})^\db{}_\da  \right) \, V_{\be\db}\,,
\nonumber\\[0.3cm]
\delta^R\,V_{\al\da} &= i\left(\theta^\ga \partial_\ga +
  \bar\theta_\dg \bar\partial^\dg\right)\, V_{\al\da} \,,\nonumber\\[0.3cm]
\delta^D\,V_{\al\da} &= \left( 3 + x^a\partial_a + \half \theta^\ga
  \partial_\ga - \half \bar \theta_\dg \bar \partial^\dg \right) \,V_{\al\da}
\,, \label{Vtransexplicit}\\[0.3cm]
\delta^S_\ga \, V_{\al\da} &= \left(-x_a \sigma^a_{\ga\dg} \,
  \delta^{\bar Q\dg} + 2 \theta_\ga \, \delta^R - i \theta^2 D_\ga -
  4i\theta_\ga  \right) \, V_{\al\da} \nonumber\\
&\quad +4i \epsilon_{\al\ga} \theta^\be \, V_{\be\da} \,,\nonumber \\[0.3cm]
\delta^{\bar S \dg} \, V_{\al\da} &= \left( x_a \sigma^{a\ga\dg} \,
\delta^Q_\ga + 2 \bar\theta^\dg \, \delta^R +i \bar\theta^2 \bar D^\dg
+4i \bar\theta^\dg \right) \, V_{\al\da} \nonumber\\
&\quad +4i\delta_\da^\dg \bar\theta^\db \, V_{\al\db} \,,\nonumber\\[0.3cm]
\delta^K_a \, V_{\al\da} &= \left(2x_a \, \delta^D + 2x^b \,
  \delta^M_{ab} -2 x_a x^b \partial_b + x^2 \partial_a + 2
  \theta\sigma_a \bar\theta \, \delta^R + \theta^2 \bar\theta^2
  \partial_a \right) \, V_{\al\da} \nonumber\\
&\quad - \theta \sigma^b \bar\theta \left( (\sigma_{ab})_\al{}^\be
  \delta_\da{}^\db + \delta_\al{}^\be (\bar\sigma_{ab})^\db{}_\da
  \right) \, V_{\be\db}\,. \nonumber
\end{align}
These results may also be obtained by working on flat space only
\cite{Pisar}.
Using (\ref{diffeom}), (\ref{Lsflach}), (\ref{oLsflach}), the transformation 
law (\ref{deltaclassV}) may be written in the form
\begin{equation}
\delta V_{\al\da} = \Lambda V_{\al\da} -\tfr{3}{2} (\sigma+ \bar \sigma)
V_{\al\da} \label{deltaVLambda}
\end{equation}
as expected from the relation between superconformal transformations and
diffeomorphisms and Weyl transformations discussed in section 3.

\section{Local Callan-Symanzik equation}

\setcounter{equation}{0}

For the quantised theory we consider for definiteness
the massless Wess-Zumino model on a
curved superspace background, which we quantise in a perturbative approach
using algebraic renormalisation  \cite{bphz}. For composite operators
we use Zimmermann's normal product algorithm  \cite{quantum} in a
version which allows for the treatment of massless fields. 
For the quantised theory,
the equation expressing the coupling of the supercurrent
to the external supergravity field is given by
\begin{gather}
\frac{\delta}{\delta H^a} \Gamma = {\ts \frac{1}{8}} \, [ V_a ] 
\cdot \Gamma \,    \label{ap}
\end{gather}
according to the action principle.
Here $\Gamma$ is the vertex functional and
the square brackets denote an insertion, a well-defined
expression for a composite operator. Denoting the dynamical
chiral field of the Wess-Zumino model by $A$, 
(\ref{ap}) yields time-ordered 1PI Green functions with an insertion of the
supercurrent by virtue of
\begin{gather}
\langle  \left[V^a (z) \right] A(z_1) \dots A(z_n)
\rangle^{1PI} = {\ts \frac{1}{8}}
\frac{\delta}{\delta A(z_n)} \dots \frac{\delta}{\delta A(z_1)}
\frac{\delta}{\delta H_a(z)} \Gamma \Big|_{H=J=\bar J=0,A=\bar
A=0}   \, .
\end{gather}
In the classical approximation, (\ref{ap}) reduces to (\ref{sch}).
Similarly we have
\begin{gather} \label{ap2}
w^{(\sigma)} \Gamma = - {\ts \frac{3}{2}} \,  [S] \cdot \Gamma \, ,
\end{gather}
which reduces to (\ref{AJswi}) in the classical approximation.
From (\ref{ap}) and (\ref{ap2}),
information about the {\it operators} $V$ and $S$ may be
extracted by performing a Legendre transform such as
to obtain $Z_c$, the generating functional for connected Green functions, 
and by applying the LSZ reduction formalism to $Z= \exp (i Z_c)$, 
the generating functional
for general Green functions.      

In \cite{ERS1}, \cite{ERS2} we have proved the validity of the {\it local}
Callan-Symanzik equation 
\begin{equation}
\Big(w\is-\gamma A\fdq{}{A}\Big)\Gamma             
+ \beta^g [\partial_g\Leff]\cdot\Gamma   = C(H,J,\bar J)\, 
\label{localCS}
\end{equation} 
to all orders in $\hbar$. Here $w\is$ is the Weyl symmetry operator
of (\ref{wis}), $A$ is the dynamical field,
and $\beta^g$ and $\gamma$ are the usual scale anomaly functions
of the massless Wess-Zumino model. $\Leff$ is an effective Lagrangian whose
form is uniquely determined by symmetry and renormalisation requirements.
We have
\begin{gather} \label{Geff}
\Geff = \int\dS \Leff + \int\dSb \Leffb \, , 
\end{gather} 
where $\Geff$ is the effective action in the sense of Zimmermann,
from which the vertex functions may be calculated.
$\Geff$ and $\Leff$ may be decomposed
into
\begin{gather}
\Geff = \Gamma_{\rm dyn} + \Gamma_{\rm geom} \, , \qquad
\Leff = \L_{\rm dyn} + \L_{\rm geom} \, , \label{geomdyn} 
\end{gather}
where $\L_{\rm dyn}$ contains terms involving the dynamical fields,
\begin{align}
\L_{\rm dyn} &= \tfr{1}{32} z \phi^3 A (\Db^2+R) e^{2iH} \bar A + \tfr{1}{48} 
\hat g
\phi^3 A^3 + \tfr{1}{8} (\xi-\epsilon) \phi^3 RA^2 + \tfr{1}{8} \epsilon \phi^3
(\Db^2+R) e^{2iH} \bar A^2 \nonumber\\
& \quad + \mbox{terms linear in $A$, $\bar A$}\,, 
\end{align}
and
$\L_{\rm geom}$ contains the purely geometrical terms
\begin{gather} \label{Lgeom}
\L_{\rm geom} =  \tfr{1}{8} \lambda_3 \Liii + \tfr{1}{16} \lambda_{R\bar R} \LRR
\,,  
\end{gather}
where
\begin{gather}
\Liii =  \phi^3 R^3 \, , \qquad
\LRR = \phi^3 (\Db^2+R) (R\bar R)  \,.
 \label{Ldef}
\end{gather}
The couplings $z$, $\hat g$, $\xi$, 
$\epsilon$, $\lambda_3$ and $\lambda_{R\bar R}$
 have been determined in \cite{ERS1},
\cite{ERS2} to all orders in $\hbar$ as power series in
the classical coupling $g$. In (\ref{Ldef}),
$R$ is the superspace curvature scalar.
The superspace curvature prepotentials are given in appendix \ref{curvature}.

The terms $C(H,J, \bar J)$ on the RHS of the local Callan-Symanzik equation
(\ref{localCS}) 
are geometrical anomalies, which according to \cite{ERS2} are given by
\begin{align}
C(H,J,\bar J) &=   
-\tfr{1}{8} (1-2\gamma) \utop
\Ltop - \tfr{1}{8} (1-2\gamma) \uW \LW \nonumber\\
&\quad + \tfr{1}{8} \left( \half \beta^g\partial_g \lambda_{R\bar R} -
  (1-2\gamma) \uRR \right) \LRR \nonumber \\
& \quad +\tfr{1}{8} \left( \half(\uiii+\uiiib) -2\gamma \uiiib \right)
(\Liii-\Liiib) \label{CC}
\, , \end{align}
with $\LRR$ as above and
\begin{align}
\Ltop &= \phi^3 \left( W^{\al\be\ga} W_{\al\be\ga} +
    (\Db^2+R)( G^a G_a - 2 R\bar R) \right) \nonumber \\[0.2cm]
\LW &= \phi^3 W^{\al\be\ga} W_{\al\be\ga} \nonumber \\[0.2cm]
\Liiib &= \phi^3 (\Db^2+R) \bar R^2\,.
\end{align}
$\Ltop$ is the chiral projection of the Gau\ss-Bonnet
density and $\LW$ is the Weyl density.
For the integrals we have
\begin{equation}
\begin{array}{lcllcl}
\Itop &=& \multicolumn{4}{l}{\int\dS \phi^3 W^{\al\be\ga} W_{\al\be\ga} +
  \int\dV E^{-1} ( G^a G_a - 2R\bar R )\,,} \\[0.2cm]
\IW &=& \int\dS \phi^3 W^{\al\be\ga}W_{\al\be\ga}\,, &
\qquad\qquad \IRR &=& \int\dV E^{-1} R\bar R \, , \\[0.2cm]
I_3 &=& \int\dV E^{-1} R^2 \,,& \qquad\qquad \bar I_3 &=& \int\dV E^{-1} \bar
  R^2\,,
\end{array} \label{geomint}
\end{equation}
where $\int\dV E^{-1}$ is the appropriate measure for a vector
superfield expression on curved superspace. 
With these expressions we have
\begin{gather}
\Gamma_{\rm geom} = \int \dS \L_{\rm geom} + \int \dSb
\Lb_{\rm geom} = {\ts \frac{1}{8}} \lambda_3 (I_3 + \bar I_3 ) 
+ {\ts \frac{1}{8}} \lambda_{R\bar R} I_{R\bar R} \, \label{Ggeom}
\end{gather}
for the geometrical contribution  to the effective
action (\ref{geomdyn}), 
which is chosen such as to eliminate those symmetry breaking terms
which are Weyl variations.
Furthermore in  an expansion around
flat superspace we have 
\begin{gather} \label{top}
\IW - \IWb = 0 \, , \qquad
\Itop + \bar\Itop = 0\, ,
\end{gather} 
which corresponds to vanishing Pontrjagin and Euler numbers.

It should be noted that the local Callan-Symanzik equation (\ref{localCS})
is chiral.
Integrating the equation over chiral superspace and adding the
complex conjugate of this integral yields the global Callan-Symanzik equation 
\begin{equation}
\C \Gamma = \Delta_{\C}^{\rg}\,, \label{globCS}
\end{equation}
where $\C$ is the Callan-Symanzik operator
\begin{gather}
\C=m\partial_m + \beta_g \partial_g  
-\gamma \N\,, \label{CSop}\\
\intertext{in which $m\partial_m$ includes all mass parameters of the theory and}
\N=\int \dS A \fdq{}{A} + \int\dSb \bar A \fdq{}{\bar A} \label{Ndef}
\end{gather}
is the counting operator for matter legs. 
The anomaly on the right hand side of (\ref{globCS}) is given by
\begin{equation}
\Delta^{\rg}_{\C} = -{\ts \frac{1}{8}} (\IW + \IWb) + {\ts \frac{1}{8}}
( \beta^g \pr_g \lambda_{R\bar R} - 2 \uRR )  \IRR 
 \, . \label{CSgeom}
\end{equation}
Here there is no contribution of $\Itop$ due to (\ref{top}).
Also at a fixed point where $\beta^g=0$, the $\IRR$ term is
absent due to Wess-Zumino consistency.

\section{Transformation properties of the supercurrent}

\setcounter{equation}{0}

\subsection{Green functions with multiple insertions}

\label{multgreensect}
The transformation properties of Green functions with multiple supercurrent
insertions are obtained by varying the quantum superconformal Ward identity
an appropriate number of times with respect to the prepotential
$H_{\al\da}$. To ensure a rigorous treatment of the Green functions with
multiple insertions generated in this way, we consider general multiple
insertion Green functions first.
For a general composite operator 
$D(x)$ coupled to an external field $h(x)$ such that
\begin{equation}
\fdq{}{h(x)} \Gamma = [D(x)]\cdot\Gamma,
\end{equation}
and a further composite operator $B(x)$ which may depend on $h(x)$,
the Zimmermann approach to composite operators yields a product rule
\begin{align}
\fdq{}{h(y)} \left( [B(x)]\cdot\Gamma\right) &= [B(x)]\cdot[D(y)]\cdot\Gamma +
  [F(x-y)]\cdot \Gamma,
\label{prodrule} \\
\mbox{with}\quad F(x-y)&=\fdq{B(y)}{h(x)}\,.
\end{align}
From general rules for functional differentiation it is clear that $F(x-y)$
is quasi-local in the sense that it contains $\delta^{(d)}(x-y)$, possibly
with some derivatives acting on it. Moreover when $B(x)=D(x)$ we find for
the second variation of $\Gamma$
\begin{align}
\fdq{}{h(x)}\fdq{}{h(y)} \Gamma &= \fdq{}{h(x)}
  \left([D(y)]\cdot\Gamma\right)
= [D(x)]\cdot [D(y)] \cdot\Gamma + [G(x-y)]\cdot\Gamma\,,
\label{DD}\\
\mbox{where}\quad G(x-y)&=\fdq{^2\Geff}{h(x)\delta h(y)}
\end{align}
is again quasi-local.
For the derivation of Ward identities it is convenient to define
\begin{align}
\{ B(x) \cdot  D(y) \} \cdot \Gamma &\equiv \left.\fdq{}{h(y)} \left(
      [B(x)]\cdot \Gamma \right) \right|_{h=0} \\
\{ D(x) \cdot D(y) \} \cdot \Gamma &\equiv \left. \fdq{}{h(x)}\fdq{}{h(y)}
      \Gamma \right|_{h=0} \,.
\end{align}
Correspondingly for triple insertions we have
\begin{align}
\{ B(x) \cdot D(y) \cdot D(z) \} \cdot \Gamma &\equiv
\left. \fdq{}{h(y)}\fdq{}{h(z)}  \left( [B(x)]\cdot\Gamma \right)
\right|_{h=0}\,, \\
\{ D(x) \cdot D(y) \cdot D(z) \} \cdot \Gamma &\equiv \left. \fdq{}{h(x)}
  \fdq{}{h(y)} \fdq{}{h(z)} \Gamma \right|_{h=0}\,,
\end{align}
for which we find according to the product rule (\ref{prodrule})
\begin{align}
\{ B(x) \cdot D(y) \cdot D(z) \} \cdot\Gamma = \phantom{+} &
\left[ \fdq{^2 B(x)}{h(y) \delta h(z)}\right] \cdot \Gamma
+ \left[ \fdq{B(x)}{h(y)}\right] \cdot [D(z)]\cdot\Gamma \nonumber \\[0.5ex]
+ &\left[ \fdq{B(x)}{h(z)}\right]\cdot [D(y)]\cdot\Gamma
+ [B(x)]\cdot[D(y)]\cdot [D(z)]\cdot\Gamma \nonumber \\[0.5ex]
+ & \,\left[ B(x)\right]\cdot[G(y-z)]\cdot\Gamma \,,
\label{bracka} \\[1ex]
\{ D(x) \cdot D(y) \cdot D(z) \} \cdot\Gamma = \phantom{+} &
\left[ \fdq{^3 \Geff}{h(x) \delta h(y)\delta h(z)}\right] \cdot \Gamma
 \nonumber \\[0.5ex]
+& \,[D(x)]\cdot [G(y-z)]\cdot\Gamma +  [D(y)]\cdot [G(z-x)]\cdot\Gamma
\nonumber \\[0.5ex]
+& \,[D(z)]\cdot [G(x-y)]\cdot\Gamma
+ [D(x)]\cdot[D(y)]\cdot[D(z)]\cdot\Gamma \,.
\label{brackb}
\end{align}
Here the first term on the right hand side of both (\ref{bracka}),
(\ref{brackb}) contains the product of two delta functions with some
derivatives acting on them.
Furthermore it is convenient to define time-ordered Green functions by
\begin{align}
\l B(x) D(y) A(z_1)\dots A(z_n)\r^{\rm 1PI} &\equiv
  \left. \fdq{^n}{A(z_1)\cdots 
  \delta A(z_n)} \{B(x)\cdot  D(y)\}\cdot\Gamma
  \right|_{A=0}\,,
\label{BDGreen}
\\
\l D(x) D(y) A(z_1)\dots A(z_n)\r^{\rm 1PI} &\equiv \left. \fdq{^n}{A(z_1)\cdots
  \delta A(z_n)} \{D(x)\cdot D(y)\} \cdot \Gamma
  \right|_{A=0} \label{DDGreen}
\end{align}
and similarly for triple insertions, with $A(z)$ a dynamical field.

Once diffeomorphism invariance is imposed, the only way to alter the
explicit expressions for $D(x)$, $D(y)$ and the quasi-local terms $F(x-y)$
in (\ref{prodrule}) and $G(x-y)$ in (\ref{DD}) is to redefine the external
field $h \to h' = h'[h]$, which amounts to a redefinition of $\Gamma[A,h]$
including its counterterms.

By introducing a source $j_A$ for $A$, the generating functional for
connected Green functions $Z_c$ is obtained by means of the Legendre
transform
\begin{equation}
Z_c[j_A, h] = \Gamma[h,A] + \int \!\!{\rm d}^d x \,\, j_A \, A \,.
\label{Legendre}
\end{equation}
Using the fact that the source $j_A$ is independent of $h(x)$, as well as
\begin{equation}
  \fdq{\Gamma[A,h]}{A(x)} = - j_A(x)\,,
\end{equation}
it is easy to show by varying (\ref{Legendre}) with respect to $h(x)$,
$h(y)$ consecutively that
\begin{align}
[D(x)]\cdot\Gamma &= [D(x)]\cdot Z_c \,, \\
[D(x)]\cdot[D(y)]\cdot\Gamma &= [D(x)]\cdot[D(y)]\cdot Z_c \,.
\end{align}
Thus it is straightforward to obtain results for connected Green functions
with double insertions once the corresponding 1PI Green functions are known.

In the subsequent we use the general results of this section for deriving
superconformal Ward identities. In particular we identify the chiral trace
term $S$ with the operator $B$, the supercurrent $V_{\al\da}$ with $D$ and
the supergravity prepotential $H_{\al\da}$ with the external field $h$.

\subsection{Single insertion}

For the quantum theory we may obtain the superconformal transformation
properties of supercurrent insertions to all orders in $\hbar$ by
proceeding in close analogy to the derivation of (\ref{Vtransclass})
in the classical approximation. We combine the local Callan-Symanzik
equation (\ref{localCS}), which is the Weyl symmetry Ward identity for
the quantum theory, with the diffeomorphism Ward identity (\ref{difW})
which is valid to all orders in $\hbar$. 
This yields 
\begin{align}
w_\al^{(\gamma)} (A,J) \Gamma &= -w_\al (H) \Gamma - \tfr{1}{8} D_\al S
\cdot \Gamma\,,
\label{anomWI} \\
\intertext{where}
w_\al^{(\gamma)}(A,J) &\equiv w_\al^\il (A,J) + w_\al^\is (A,J)
-\tfr{1}{12} \gamma \, D_\al \left(A \textstyle \fdq{}{A} \right) \,, \\[1ex]
S &\equiv \tfr{2}{3} [\beta^g \partial_g \Leff] - \tfr{2}{3} C(H, J, \bar
J) \,.
\end{align}
The corresponding antichiral Ward identity may be obtained from
(\ref{anomWI}) by complex conjugation; so far we have not imposed $\bar
D^\da \Omega^\al = D^\al \bar \Omega^\da$. From (\ref{anomWI}) we may
obtain Ward identities for Green functions with arbitrarily many
supercurrent insertions by varying with respect to $H^{\al\da}$ the
required number of times. Varying once with respect to $H^{\al\da}$, and
subsequently restricting to flat space, we obtain for a single supercurrent
insertion the Ward identity
\begin{align}
w_\al^{(\gamma)}(A) \left( [V_{\be\db}(z')]\cdot\Gamma \right) = &
-\tfr{1}{16} \bar D^\da \left( \{ V_{\al\da}(z) \cdot V_{\be\db}(z')\}
  \cdot \Gamma \right) 
 + \Delta_{\al\be\db}{}^{\ga\dg}(z,z') \left( [V_{\ga\dg}(z')]\cdot\Gamma
\right) \nonumber \\[0.5ex]
& -\tfr{1}{12} D_\al \{ \beta^g \partial_g \Leff(z) \cdot V_{\be\db}(z')\}
\cdot \Gamma \,,
\label{localWIV}
\end{align}
where according to the general discussion of the preceding section
\ref{multgreensect} we have
defined
\begin{align}
\{ V_{\al\da}(z) \cdot V_{\be\db}(z')\} \cdot \Gamma &\equiv
 64 \left. \fdq{^2}{H^{\al\da}(z)\delta H^{\be\db}(z')} \Gamma
 \right|_{H=J=0} \nonumber \\[0.5ex]
& = [V_{\al\da}(z) ] \cdot [V_{\be\db}(z')]\cdot \Gamma +
 [G_{\al\da\be\db}(z-z')] \cdot\Gamma \,,
\label{VVins}\\[1ex]
\mbox{where} \quad G_{\al\da\be\db}(z-z') &\equiv 64 \fdq{^2
  \Geff}{H^{\al\da}(z) \delta H^{\be\db}(z')}\,, \quad \mbox{and}
\label{Gab} \\
\{ \beta^g \partial_g \Leff(z) \cdot V_{\be\db}(z') \} \cdot \Gamma &\equiv
8 \left. \fdq{}{H^{\be\db}(z')} \left( [\beta^g\partial_g
 \Leff(z)]\cdot\Gamma\right) \right|_{H=J=0} \nonumber \\[0.5ex]
&= [\beta^g\partial_g\Leff(z)] \cdot [V_{\be\db}(z')]\cdot\Gamma + 8 \left[
 \fdq{(\beta^g\partial_g\Leff(z))}{H^{\be\db}(z')}\right] \cdot\Gamma \,.
\label{LVins}\\
\intertext{Furthermore for convenience we have defined }
\Delta_{\al\be\db}{}^{\ga\dg}(z,z') \, \big( \bullet \big) &\equiv 
-\tfr{1}{4} \delta_\be^\ga \, \delta_\db^\dg \, \bar D^\da \big (\{ D_\al, \bar
D_\da \} \delta^8(z-z') \,\bullet\, \big) \nonumber \\
&\quad -\tfr{1}{4} \delta_\al^\ga \, \{ D_\be \bar D_\db\} \bar D^\dg
\big (\delta^8(z-z') \,\bullet \, \big ) \nonumber \\ 
&\quad -\tfr{1}{4} \delta_\be^\ga \, \delta_\db^\dg \, \bar D^2 \big (D_\al
\delta^8(z-z')  \,\bullet\,\big ) \,,
\label{Deltadef}
\end{align}
such that
\begin{equation}
-8 \left. \fdq{}{H^{\be\db}(z')}\left( w_\al^\il(H)\right) \Geff
  \right|_{H=0} = \Delta_{\al\be\db}{}^{\ga\dg}(z,z') V_{\ga\dg}(z)
\end{equation}
expresses the chiral part of the conformal transformation properties of the
supercurrent. It should be noted that the geometrical terms $C(H,J,\bar J)$
do not contribute to (\ref{localWIV}) since they are of second order in $H$.

The full quantum superconformal transformation properties of the
supercurrent insertion may be obtained by multiplying (\ref{localWIV}) by
$\Omega^\al$, integrating, adding the complex conjugate and imposing $\bar
D^\da \Omega^\al = D^\al \bar \Omega^\da$. In this case the $\{V_{\al\da}
\cdot V_{\be\db}\}$ terms drop out and we obtain
\begin{align}
W^{(\gamma)} [V_{\al\da}(z')]\cdot \Gamma &= [\delta V_{\al\da}(z')] \cdot
\Gamma - \int\dS \sigma(z) \{ \beta^g \partial_g \Leff(z) \cdot
V_{\al\da}(z') \} \cdot \Gamma + c.c. \,, 
\label{Vquanttrans}\\
\intertext{where}
W^{(\gamma)} &\equiv \int \dV \Omega^\al w_\al^{(\gamma)}(A) + c.c.
\label{Wgammadef}
\end{align}
is the Ward operator for the dynamical fields.
$\delta V_{\al\da}$ is the classical conformal transformation of the
supercurrent as given by (\ref{deltaclassV}), (\ref{deltaVLambda}). In the
classical approximation, (\ref{Vquanttrans}) reduces to (\ref{Vtransclass})
again with no symmetry breaking terms present.
Since $\delta V_{\al\da}$ in (\ref{Vquanttrans}) is identical to the
expression which appears in the classical approximation
(\ref{deltaVLambda}), the supercurrent does not acquire any anomalous
dimension in quantum theory. This is in agreement with the fact that  the
supercurrent contains the supersymmetry currents and the energy-momentum
tensor among its components. These currents are conserved and hence do not
have anomalous dimensions.

According to (\ref{sigmacoeff}), only $R$ transformation, dilatation,
 special supersymmetry 
 and special conformal transformations  are anomalous, but not the
 super Poincar\'e transformation.
For dilatations for which $\sigma$ is a real constant,
(\ref{Vquanttrans}) simplifies to 
\begin{equation}
\left(- W^D + \beta^g\partial_g - \gamma \N \right) \left(
  [V_{\al\da}] \cdot \Gamma  \right) + [\delta^{D}V_{\al\da}] \cdot
  \Gamma =0 \label{VDil}
\end{equation}
by virtue of (\ref{Geff}) and with $\N$ as in (\ref{Ndef}),
$\delta^DV_{\al\da}$ as in (\ref{Vtransexplicit}).
Here the dilatation Ward operator is given by
\begin{align}
 W^D &=  \int\dV \Omega^\al (w^\il_\al(A) + w^\is_\al(A)) +
 c.c. \, , \\
\Omega^{\al D} &= \half \theta^\al \bar\theta^2 -\tfr{i}{2} x^a
 \sigma_a^{\al\da} \bar\theta_\da \, . \nonumber 
\end{align}

Using the general results of section \ref{multgreensect} it is
straightforward to find the superconformal Ward identities for Green
functions which follow from (\ref{Vquanttrans}). Defining
\begin{align}
\l  V_{\al\da}(z_A) \, X \r &\equiv 8
\left(\tfr{\hbar}{i}\right)^{n+m+1}
\left.  \fdq{^{n+m}}{j_A^n \delta \bar j_A^m}
\fdq{}{H^{\al\da}(z_A)} Z \right|_{j=\bar j=H=J=0} \,, 
\label{VVXGreen}\\
\l S(z_A) V_{\al\da}(z_B) \, X \r &\equiv  8
\left(\tfr{\hbar}{i}\right)^{n+m+1}
\left. \fdq{}{H^{\al\da}(z_B)} \left( [S(z_A)]\cdot Z\right)\right|_{j=\bar
  j=H=J=0}\,, 
\label{SVXGreen}
\end{align}
with $X\equiv A(z_1) \dots A(z_n) \bar A(z_1') \dots \bar A(z_m')$ 
and $Z \equiv \exp (\tfr{i}{\hbar} Z_c)$, and performing the Legendre
transform of (\ref{Vquanttrans}) we find
\begin{equation}
\delta \l V_{\al\da}(z) \, X\r = \int\dS\!' \sigma(z') \, \l 
\beta_g \partial_g \, \Leff(z')  V_{\al\da}(z) \, X \r + c.c. \,.
\label{VtransGreen}
\end{equation}
Here the conformal transformation $\delta$ of the Green function is given
by
\begin{align}
\delta \, \l V_{\al\da}(z) \, X \r = \l \delta V_{\al\da}(z) \,
X \r 
& + \sum_{k=1}^n \l V_{\al\da}(z) \, A(z_1) \dots \delta A(z_k) \dots
  A(z_n) \, \bar A(z_1') \dots \bar A(z_M')\r \nonumber \\
& + \sum_{k=1}^m \l V_{\al\da}(z) \, A(z_1) \dots
  A(z_n) \, \bar A(z_1') \dots \delta \bar A(z_k') \dots \bar A(z_M')\r \,, 
\label{deltadef}
\end{align}
with $\delta V_{\al\da}(z)$ given by (\ref{deltaclassV}),
(\ref{deltaVLambda}) and 
\begin{equation}
\delta A = \left( \Lambda -(1+\gamma) \, \sigma \right) \, A \,, \quad
\delta \bar A =  \left( \bar \Lambda -(1+\gamma) \, \bar \sigma \right) \,
\bar A \,.
\end{equation}
(\ref{VtransGreen}) is a concise expression for the superconformal Ward
identity expressing the breakdown of superconformal symmetry at the level
of Green functions.
The fact that the presence of the supercurrent insertion does not generate
further anomalies is due to supersymmetry. In particular, there are no
further anomalies since the auxiliary mass terms arising in the
renormalisation process depend on $J$, $\bar J$ but are independent of
$H_{\al\da}$ which couples to the supercurrent.

\subsection{Double insertion}

Varying the Ward identity (\ref{anomWI}) twice with respect to $H^{\be
  \db}(z_1)$, 
$H^{\ga \dg}(z_2)$ we obtain the trans\-forma\-tion proper\-ties of
Green functions with double supercurrent insertions.
In this context the two point function $\l V_{\al\da}(z_1)
V_{\be\db}(z_2)\r$ is of special interest since there are potential
contributions from the geometrical terms $C(H,J,\bar J)$.
The second variation of (\ref{anomWI}) yields when restricting to flat
space 
\begin{align}
w_\al^{(\gamma)} (A(z)) \left( \{V_{\be\db}(z_1) \cdot V_{\ga\dg}(z_2)\} \cdot
  \Gamma \right) = &
-\tfr{1}{16} \bar D^\da \{ V_{\al\da}(z) \cdot V_{\be\db}(z_1) \cdot
  V_{\ga\dg}(z_2) \} \cdot \Gamma   \nonumber \\[0.5ex]
& + \Delta_{\al\be\db}{}^{\de\dd} (z, z_1) \{ V_{\de\dd}(z) \cdot
  V_{\ga\dg}(z_2) \} \cdot \Gamma   \nonumber \\[0.5ex]
& + \Delta_{\al\ga\dg}{}^{\de\dd} (z,z_2) \{ V_{\de\dd}(z) \cdot
  V_{\be\db}(z_1) \} \cdot \Gamma   \nonumber \\[0.5ex]
& + \Delta^{(2)}_{\al\be\db\ga\dg}{}^{\de\dd}(z; z_1,z_2)
  [V_{\de\dd}(z)]\cdot \Gamma       \nonumber \\[0.5ex]
& - \tfr{1}{12} D_\al \{ \beta^g \partial_g \Leff(z)  \cdot V_{\be\db}(z_1) 
  \cdot V_{\ga\dg}(z_2)\} \cdot \Gamma  \nonumber \\[0.5ex]
& -\tfr{1}{12} D_\al \A_{\be\db\ga\dg}(z-z_1, z-z_2) \,.
\label{WIdouble}
\end{align}

Here the triple insertions 
\begin{align}
\{ V_{\al\da}(z) \cdot V_{\be\db}(z_1) \cdot V_{\ga\dg}(z_2) \} &\equiv 8^3
\left. \fdq{^3}{H^{\al\da}(z) \delta H^{\be\db}(z_1) \delta
    H^{\ga\dg}(z_2)}  \Gamma \right|_{H=J=0} \,,
\label{triple} \\
\{ \beta^g\partial_g \Leff(z) \cdot V_{\be\db}(z_1) \cdot V_{\ga\dg}(z_2)
\} &\equiv 64 \left. \fdq{^2}{H^{\be\db}(z_1) \delta H^{\ga\dg}(z_2)}
  \left( [\beta^g\partial_g\Leff] \cdot \Gamma \right) \right|_{H=J=0}
\end{align}
may be evaluated according to the general rules (\ref{bracka}),
(\ref{brackb}). 
$\Delta_{\al\be\db}{}^{\ga\dg}$ is given by (\ref{Deltadef}), and
$\Delta^{(2)}_{\al\be\db\ga\dg}{}^{\de\dd}(z;z_1,z_2)$ is defined such that
\begin{equation}
\Delta^{(2)}_{\al\be\db\ga\dg}{}^{\de\dd}(z;z_1,z_2) V_{\de\dd}(z) = 
-64 \fdq{}{H^{\be\db}(z_1)} \fdq{}{H^{\ga\dg}(z_2)} \left( w_\al^{(2)}
  (H(z)) \right) \Gamma \,;
\label{Delta2}
\end{equation}
the explicit expression may be found in appendix \ref{appdoubleins}.
$\A_{\be\db\ga\dg}(z-z_1, z-z_2)$ denotes the contributions from the purely
geometrical terms,
\begin{equation}
\A_{\be\db\ga\dg}(z-z_1, z-z_2) \equiv 64 \left. \fdq{}{H^{\be\db}(z_1)}
\fdq{}{H^{\ga\dg}(z_2)} C(H, J, \bar J) (z) \right|_{J=J=0}\,.
\end{equation}
According to the explicit expression for $C(H,J,\bar J)$ given by (\ref{CC})
we decompose the form $\A_{\al\da\be\db}(z-z_1,z-z_2)$ into
\begin{align}
\A_{\al\da\be\db}(z-z_1,z-z_2) &= -8 (1-2\gamma) \utop \A^{\rm
  top}_{\al\da\be\db}(z-z_1,z-z_2) \nonumber \\
&\quad - 8 (1-2\gamma) \uW \A^{\rm
  Weyl}_{\al\da\be\db} (z-z_1,z-z_2) \nonumber \\
&\quad +8 \left( \half \beta^g\partial_g \lambda_{R\bar R} -
  (1-2\gamma) \uRR \right) \A^{R\bar R}_{\al\da\be\db}(z-z_1,z-z_2) \nonumber \\
&\quad + 8 \left( \half (\uiii+\uiiib) -2\gamma \uiiib \right)
  \A^3_{\al\da\be\db}(z-z_1,z-z_2)  \,. 
\label{AA}
\end{align}
For the topological contribution we find
\begin{align}
\A^{\rm top}_{\al\da\be\db}(z-z_1,z-z_2) =
\bar D^2 & \Biggl[ \,\,\,
\pr_a K^a{}_{\al\da\be\db}(z-z_1, z-z_2)  \nonumber \\
& + D^2 L_{\al\da\be\db}(z-z_1, z-z_2)\label{Atop}
 \\[0.5ex]& 
+ \pr_a D^\epsilon M^a{}_{\epsilon,\al\da\be\db} (z-z_1, z-z_2)  \nonumber \\
& - \pr_{\al\da} \pr_{\ga\dg} N^{\ga\dg}{}_{\be\db}(z-z_1, z-z_2) 
- \pr_{\be\db} \pr_{\ga\dg} N^{\ga\dg}{}_{\al\da}(z-z_2, z-z_1)
\, \Biggr] \,,
\nonumber 
\end{align}
where we have used the linearised expressions of appendix \ref{curvature}
for the curvature tensors.
Here
\begin{align}
K^a{}_{\al\da\be\db}(z-z_1, z-z_2) &= 
 8 \delta^8(z-z_1) \pr^a \delta G_{\al\da\be\db}(z-z_2) + 4i
\delta^8(z-z_1) \sigma^a_{\ga\dg} D^\ga \bar D^\dg \delta
G_{\al\da\be\db}(z-z_2) \nonumber \\
& \quad -4i \epsilon^{abcd} \sigma_{c\al\da} \sigma_d^{\ga\dg}
\delta^8(z-z_1) \pr_b 
\delta G_{\ga\dg\be\db}(z-z_2) \nonumber \\
&  \quad -\tfr{2}{3} i \sigma^a_{\ga\dg} \delta^8(z-z_1) [D_\al, \bar D_\da]
\delta 
G^{\ga\dg}{}_{\be\db} (z-z_2) \nonumber \\
& \quad + \left((\al\da, z-z_1) \leftrightarrow (\be\db, z-z_2)\right) \,,
\label{Kdef}
\end{align}
where we have defined
\begin{equation}
\delta G_{\al\da\be\db}(z-z') \equiv \left. \fdq{}{H^{\be\db}(z')}
  G_{\al\da}(z) \right|_{H=J=0} \,,
\end{equation}
with $G_{\al\da}$ the real curvature tensor.
The explicit expressions for $L$, $M$, $N$ in (\ref{Atop}) are not relevant
for our subsequent calculations and are listed in appendix
\ref{appdoubleins} for completeness. From (\ref{Atop}) it is easy to see
that $\A^{\rm top}$ vanishes when integrated over ${\rm d}^6z$, as is
expected for a topological density expanded around flat space.

Furthermore we have for the remaining forms in (\ref{AA})
\begin{subequations}
\begin{align}
\A^{\rm Weyl}_{\al \da \be \db} (z-z_1, z-z_2) & = - 8 \,
\E^{\gamma \delta}{}_{\al},^{\eps \eta}{}_\be \bar D^2 \pr_{\ga \da} D_{\de}
\delta^8(z-z_1) \bar D^2 \pr_{\de\db} D_\eta \delta^8(z-z_2)
\, , \label{AWeyl}\\ 
\A^{R \bar R}_{\al \da \be \db} (z-z_1, z-z_2)  & = \tfr{4}{9}
\bar D^2 \Bigl( \bar D^2 \pr_{\al\da} \delta^8(z-z_1) D^2 \pr_{\be\db}
\delta^8(z-z_2)
\nonumber \\
& \qquad\qquad  + \bar D^2 \pr_{\be\db} \delta^8(z-z_2) D^2 \pr_{\al\da}
\delta^8(z-z_1) \Bigr)
\,,\label{ARR}\\
\A^3_{\al\da\be\db}(z-z_1,z-z_2) &= \tfr{8}{9} \bar D^2 \left( D^2 \pr_{\al\da}
  \delta^8(z-z_1) D^2 \pr_{\be\db} \delta^8(z-z_2) \right)
\,, \label{A3}
\end{align}
\end{subequations}
where we have used $\pr_{\al\da} \equiv \sigma^a_{\al\da} \pr_a$ and
$\E$ is the totally symmetric symbol
\begin{equation}
\E_{\al \be \ga},^{\de \eps \eta}  = \delta_{\al}{}^{( \de} 
\delta_{\be}{}^{\eps} \delta_{\ga}{}^{\eta)} \, .
\end{equation}
As before we obtain the superconformal transformation properties of the
double insertion by multiplying the Ward identity (\ref{WIdouble}) by
$\Omega^\al$ , integrating, adding the complex conjugate and imposing $\bar
D^\da \Omega^\al=D^\al \bar \Omega^\da$. 
In this case the $\{ V_{\al\da} \cdot V_{\be\db} \cdot V_{\ga\dg} \}$ terms
drop out and we have
\begin{align}
W^{(\gamma)}(A) \{ V_{\be\db}(z_1) \cdot V_{\ga\dg}(z_2) \} \cdot \Gamma =&
\phantom{+} \{ \delta V_{\be\db}(z_1) \cdot V_{\ga\dg}(z_2) \} \cdot \Gamma
+ \{ V_{\be\db}(z_1) \cdot \delta V_{\ga\dg}(z_2) \} \cdot \Gamma
\nonumber \\[0.5ex]
&+ \chi_{\be\db}(z_1,z_2) [V_{\ga\dg}(z_1)] \cdot \Gamma
+ \chi_{\ga\dg}(z_2,z_1) [V_{\be\db}(z_2)]\cdot\Gamma \nonumber \\[0.5ex]
&- \int \dS \sigma(z) \{ \beta^g\pr_g\Leff(z) \cdot V_{\be\db}(z_1) \cdot
V_{\ga\dg}(z_2) \} \cdot \Gamma + c.c. \nonumber \\[0.5ex]
&- \int \dS \sigma(z) \A_{\be\db\ga\dg}(z-z_1, z-z_2) + c.c. \,,
\label{WIdoubleint}
\end{align}
with $W^{(\ga)}$ as in (\ref{Wgammadef}) and $\chi_{\be\db}$ is obtained
from integrating the $\Delta^{(2)}$ terms of (\ref{Delta2}),
\begin{gather}
\chi_{\be\db}(z_1,z_2) V_{\ga\dg}(z_1) + \chi_{\ga\dg}(z_2,z_1)
V_{\be\db}(z_2) \equiv \int\dV \Omega^\al(z)
\Delta^{(2)}_{\al\be\db\ga\dg}{}^{\de\dd} V_{\de\dd}(z) + c.c. \,, \\[0.5ex]
\chi_{\al\da}(z_1,z_2) = \{ D_\al,\bar D_\da\} \bar D^2 \Omega^\be(z_1)
D_{1\be} \delta^8(z_1-z_2) + c.c.\,.
\end{gather}

Let us now investigate the 
consequences of the Ward identities (\ref{WIdouble})
and (\ref{WIdoubleint}) for Green functions. These are defined according to
the general rules 
(\ref{BDGreen}) and (\ref{DDGreen}) and in analogy to (\ref{VVXGreen}),
(\ref{SVXGreen}). For Green functions involving also dynamical fields in
addition to the insertions we find from the integrated Ward identity
(\ref{WIdoubleint}) 
\begin{align}
\delta \, \l V_{\al\da}(z_A) V_{\be\db}(z_B) X \r &=
 \int \dS \sigma(z) \, \l \beta^g\pr_g \Leff(z)  V_{\al\da}(z_A)
V_{\be\db}(z_B) \, X \r + c.c. \nonumber \\
& \quad - \left( \chi_{\al\da}(z_A, z_B) \l V_{\be\db} (z_A) X \r
  + \chi_{\be\db} (z_B, z_A) \l V_{\al\da}(z_B) X \r \right) \,,
\end{align}
where $X \equiv A(z_1) \dots A(z_n) \bar A(z_1') \dots
\bar A(z_m')$ and the conformal transformation $\delta$ is defined in
analogy to (\ref{deltadef}).
For Green functions involving dynamical fields there are no contributions
from the geometrical terms. However there are such contributions for the
two point function 
\begin{equation} 
\l V_{\al\da} (z_1) V_{\be\db}(z_2) \r \equiv \left .\{
  V_{\al\da}(z_1) \cdot V_{\be\db}(z_2) \} \cdot Z \right|_{A=\bar A=0}
= 64 \frac{\delta^2}{\delta H^{\al \da} (z_1) \delta H^{\be \db}(z_2)} Z
\Big|_{H=J=A=\bar A=0}  \,, 
\label{VVvac}
\end{equation}
which we consider now in detail.
In this case the non-integrated Ward identity (\ref{WIdouble}) 
involves also the three point function which according to (\ref{brackb}),
(\ref{triple}) is
given by
\begin{align} \label{3ptfn}
\l V_{\al\da}(z) \cdot V_{\be\db}(z_1) \cdot V_{\ga\dg}(z_2)  \r &= 8^3
\frac{\delta^3}{\delta H^{\al \da} (z) \delta H^{\be \db}(z_1) \delta
H^{\ga \dg}(z_2)} Z \Big|_{H=J=A=\bar A=0} \\
& = \l [V_{\al\da}(z)] \cdot [V_{\be\db}(z_1)] \cdot [V_{\ga\dg}(z_2)]  \r
\nonumber\\ & \;\; + 
\l [V_{\al\da}(z)] \cdot [G_{\be\db\ga\dg}(z_1-z_2)]  \r \; + \; {\rm
permutations} \,  \nonumber \\ & \;\; + F_{\al \da \be \db \ga \dg}(z,z_1,z_2)
\, , \nonumber \end{align}
with $G_{\be\db\ga\dg}$ as in (\ref{Gab}) and
\begin{gather}
F_{\al \da \be \db \ga \dg}(z,z_1,z_2) \equiv 8^3 
\frac{\delta^3}{\delta H^{\al \da} (z) \delta H^{\be \db}(z_1) \delta
H^{\ga \dg}(z_2)} \Gamma_{\rm geom} \Big|_{H=J=0} \, ,
\end{gather}
where $\Gamma_{\rm geom}$ is given by (\ref{Ggeom}). $F$ contains 
terms involving the product of two delta functions in agreement
with (\ref{brackb}) and reflects contributions of
counterterms in the action which have been added to eliminate
those symmetry breaking geometrical terms which are variations.
The expression (\ref{3ptfn}) 
for the three point function is clearly symmetric.

For (\ref{VVvac}), (\ref{3ptfn}) the Ward identity (\ref{WIdouble}) reads
\begin{align}
\tfr{1}{16} \bar D^\da \l V_{\al\da}(z) V_{\be\db}(z_1)
V_{\ga\dg}(z_2) \r &=
\phantom{+} \Delta_{\al\be\db}{}^{\de\dd} (z,z_1) \l V_{\de\dd}(z)
V_{\ga\dg}(z_2) \r \nonumber \\
&\quad + \Delta_{\al\ga\dg}{}^{\de\dd}(z,z_2) \l V_{\be\db}(z_1)
V_{\de\dd}(z) \r \nonumber \\
& \quad - \tfr{1}{12} D_\al \l \beta^g\pr_g\Leff(z) \, V_{\be\db}(z_1)
V_{\ga\dg}(z_2) \r \nonumber \\
& \quad -\tfr{1}{12} D_\al \A_{\be\db\ga\dg}(z-z_1, z-z_2) \,,
\label{WIdoublevac}
\end{align}
where we have used that $\l V_{\al\da}(z_1)\r$ vanishes since the
corresponding diagram is a tadpole with degree of divergence $\delta_V=3$.
The $\Delta$ terms are given by (\ref{Deltadef}).

The Ward identity (\ref{WIdoublevac}) relates the supercurrent two
and three point functions. Its form is essentially unique except for
the explicit expressions for the quasi-local terms $G$, $F$ in (\ref{3ptfn})
and $\Delta$ 
in (\ref{WIdoublevac}), which are determined by the
approach chosen for introducing curved superspace and in particular
by our definitions for the prepotentials $H_{\al \da}$ and $J$. 
Due to our approach to deriving Ward identities on curved superspace,
the quasi-local terms  $\Delta$ in
(\ref{WIdoublevac}) encode the superconformal transformation properties of the
supercurrent and lead to (\ref{deltaclassV}) and (\ref{deltaVLambda}) upon
integration. 
At the same time our choice for $H_{\al \da}$ determines the explicit form
for the quasi-local terms $G$, $F$ in (\ref{3ptfn}), as was
discussed  in section \ref{multgreensect}.

It is perhaps instructive to compare the Ward identity (\ref{WIdoublevac})
with the Ward identity discussed by
Osborn in \cite{Osborn} (equation (7.41) there)
for $\beta^g=0$ and discarding the
geo\-metri\-cal terms. The Ward identity (7.41) 
of \cite{Osborn} agrees with (\ref{WIdoublevac}) above up to a shift
between the quasi-local terms.
The terms involving
$\Delta$ in (\ref{WIdoublevac}) correspond to the
terms on the right hand side of the Ward identity
(7.41) in \cite{Osborn}. However the
explicit expressions for these terms differ in the two approaches, which is
compensated by different choices for the quasi-local terms $G$
in (\ref{3ptfn}), whose definition is given in (\ref{Gab}).
In \cite{Osborn} the $G$ terms are chosen such as to obtain the
simplest form for the terms in the Ward identity (7.41) which corresponds to
$\chi_{\al\da\be\db}{}^{\ga\dg}=0$ in the notation used there. 
Since the shift between quasi-local terms corresponds a
redefinition of the prepotential, $H_{\al \da} \rightarrow
\tilde H_{\al \da} [H]$, 
it would be interesting to see if a redefinition exists
which transforms our approach and the
 approach of \cite{Osborn}  into each other. However direct
comparison is difficult since in the two cases a different representation
is used for the diffeomorphisms such that $\Lambda$ defined in
(\ref{diffeom}) of this paper is different from $h$ defined in (2.3) of
\cite{Osborn}. 

Moreover any redefinition of $H_{\al \da}$  as a function of itself
leaves the flat space supercurrent
transformation laws (\ref{Vtransexplicit}) invariant in the superconformal
case when $\bar D^\da \Omega^\al=D^\al \bar \Omega^\da$. 
Non-trivial redefinitions  are at least  of second order,
$\tilde H_{\al \da} = H_{\al \da} + O(H^2)$, from which follows that the
only change of the Ward identity (\ref{WIdoublevac}) induced by the
redefinition is a shift between the $\Delta$ terms in (\ref{WIdoublevac})
and the $G$ terms in (\ref{VVins}). This is equivalent to the statement
in (7.39) of \cite{Osborn} that the flat space supercurrent transformations
are unique up to terms involving 
$(\bar D^\da \Omega^\al - D^\al \bar \Omega^\da)$.

The integrated Ward identity (\ref{WIdoubleint})
for the two point function (\ref{VVvac}) reads
\begin{align}
\delta \, \l V_{\al\da}(z_1) V_{\be\db}(z_2) \r &=  \int\dS \sigma(z)
\, \l  \beta^g\pr_g \Leff(z) \, V_{\al\da}(z_1) V_{\be\db}(z_2) \r +
c.c. \nonumber \\
&\quad - 8 \int\dS \sigma(z) \Bigl( (1-2\gamma) \uW \A^{\rm
    Weyl}_{\al\da\be\db} (z-z_1, z-z_2) \nonumber \\
&\quad \qquad
- (\half \beta^g\pr_g
  \lambda_{R\bar R} - (1-2\ga) \uRR) \A^{R\bar R}_{\al\da\be\db}(z-z_1,
  z-z_2) \Bigr) + c.c. \, .
\label{WIdoubleintvac}
\end{align}
Here the three point function present in (\ref{WIdoublevac})
has dropped out upon integrating, adding the complex conjugate
and imposing $\bar D^\da \Omega^\al = D^\al \bar \Omega^\da$.
Furthermore as far as the purely geometrical terms are concerned, it is
easy to see 
from (\ref{Atop}), (\ref{A3}) 
that neither $\A^{\rm top}$
nor $\A^3$ contribute to (\ref{WIdoubleintvac}) since
\begin{equation}
\int\dS \sigma(z) \A^{\rm top}_{\al\da\be\db}(z-z_1,z-z_2) + \int\dSb
\bar\sigma(z) \bar \A^{\rm top}_{\al\da\be\db}(z-z_1, z-z_2) =0
\label{noAtop}
\end{equation}
and similarly for $\A^3$. This is due to the fact that for superconformal
transformations, 
$\sigma$ and $\bar\sigma$ as given by (\ref{sigmacoeff}) are restricted by
\begin{subequations}
\begin{gather}
DD\sigma=0\,, \qquad D_\al \pr_a \sigma=0\,, 
\label{sigmarestricta}\\
\pr_a\sigma=\pr_a\bar\sigma\,.
\label{sigmarestrictb}
\end{gather}
\end{subequations}
It is important to note that only the combination of the chiral and antichiral
integrals in (\ref{noAtop}) vanishes, not the terms involving $\sigma$ or
$\bar\sigma$ separately. When inserting the lengthy expression (\ref{Atop})
into (\ref{noAtop}), the terms involving $\pr_a$ given by $K$ in
(\ref{Kdef}) cancel using
(\ref{sigmarestrictb}), 
while the terms involving $D^2$, $\pr_aD_\be$ or $\pr_a\pr_b$ vanish due to (\ref{sigmarestricta}).

At a renormalisation group fixed point where $\beta^g=0$,
(\ref{WIdoubleintvac}) is 
simplified further. However when discussing fixed points
we have to bear in mind that within our
perturbative approach to the Wess-Zumino model
we may only reach the trivial fixed point where
$\beta^g=\gamma=g=0$.
The  three point function on the right hand side of (\ref{WIdoubleintvac})
vanish in this case. 
Moreover, as shown in \cite{ERS2}, consistency implies that the
coefficient of $\A^{R\bar R}$ in (\ref{AA}) vanishes at fixed points.
Thus we obtain
\begin{equation}
\left. \delta \left\l V_{\al\da}(z_1)\,V_{\be\db}(z_2)\right\r
\right|_{\beta^g=0}
 = -8  \uW \int\dS \sigma(z) \A^{\rm
    Weyl}_{\al\da\be\db}(z-z_1,z-z_2) +c.c. \,. 
\end{equation}
Therefore at the fixed point, the conformal transformation properties of the
supercurrent two point function are entirely determined by the Weyl
anomaly.
 This result is the supersymmetric generalisation of a similar result
obtained for the energy-momentum tensor two point function for
non-supersymmetric theories in \cite{EO}. Furthermore it is in 
agreement with the
fact that for superconformal theories, the supercurrent two point function
$\l V_{\al\da}(z_1)V_{\be\db}(z_2)\r$ itself is entirely determined by the
Weyl anomaly as was shown in \cite{Osborn}.

For scale transformations we may obtain an equation analogous to
(\ref{VDil}) for double insertions. 
For $\sigma=-1$, (\ref{WIdoubleint}) reduces to
\begin{multline}
\left( - W^D + \beta^g\partial_g -\gamma\N \right) \{V_{\al\da}(z_1)
  \cdot    V_{\be\db}(z_2)\} \cdot \Gamma 
+ \delta^{\rm D} \{V_{\al\da}(z_1)\cdot V_{\be\db}(z_2)\} \cdot\Gamma \\
=\, - \int\dS  \A_{\al\da\be\db}(z-z_1,z-z_2)+c.c. \,,
\label{WIdil}
\end{multline}
with $\delta^{\rm D}V_{\al\da}$ given by (\ref{Vtransexplicit}) and
$\delta^D \{ V_{\al\da}\cdot V_{\be\db}\} = \{ \delta^D V_{\al\da} \cdot
  V_{\be\db} \} + \{ V_{\al\da} \cdot \delta^D V_{\be\db}\}$. 
Only $\A^{\rm Weyl}$ and $\A^{R\bar R}$ contribute to $\A$ in (\ref{WIdil}).
For the two point function (\ref{WIdil}) yields 
\begin{equation}
\delta^{\rm D} \l  V_{\al\da}(z_1) V_{\be\db}(z_2) \r
+ \beta^g \partial_g \l V_{\al\da}(z_1) V_{\be\db}(z_2) \r
= -\int\dS  \A_{\al\da\be\db}(z-z_1,z-z_2) +c.c. \,.
\end{equation}
Similarly we find for the $R$ transformations
\begin{equation}
\delta^R \l  V_{\al\da}(z_1) V_{\be\db}(z_2) \r
- \tfr{2}{3} i\l \beta^g \partial_g \left( \int \dS \Leff(z) - \int\dSb
  \bar \Leff (z) \right) 
 V_{\al\da}(z_1) V_{\be\db}(z_2) \r
= 0  \,.
\end{equation}
Here the geometrical terms drop out due to (\ref{top}), (\ref{ARR}),
(\ref{A3}). A similar relation holds for Green functions involving also
dynamical fields in addition to the supercurrent insertions.

\section{Supercurrent Components}
\setcounter{equation}{0}

In this section we describe in detail how the energy-momentum tensor, 
the supersymmetry current and the $R$ current 
are related to the components of the supercurrent. In parti\-cular we 
show that the energy-momentum tensor
is coupled to the metric as expected. Furthermore we
discuss improvement terms. The discussion of this section allows for
the comparison with a related approach which uses the component
formalism \cite{PW}.

\subsection{Definition of component currents}

From the flat space definition of the energy-momentum tensor according to
which it is the conserved current associated with translation invariance, 
we expect to find the divergence of $T_{a b}$ in a local Ward identity 
whose parameter is the translation component of $\Omega^\al$ in  
(\ref{omegasol}), with $t^a=t^a(x)$. However there is an arbitrariness
in generalising the component $\Omega^\al_P$ in  
(\ref{omegasol}) to a function $\Omega^{\al}_P[t^a(x)]$ 
which reduces to (\ref{omegasol}) again in the limit $t^a=const.$
In fact there is a whole equivalence class of functions 
$\Omega^{\al}_P[t^a(x)]$, the elements of which differ by derivatives of
$t^a(x)$.  We make use of this  arbitrariness to choose
$\Omega^{\al}_P[t^a(x)]$ such that the corresponding Weyl parameter
$\sigma$ vanishes,
\begin{equation}
\Omega^{\al}_P[t^a(x)] = -\tfr{i}{2}\sigma_a^{\al\da} t^a(x)
-\tfr{1}{4} \theta^\al\bar\theta^2 \partial^c t_c(x) \,.
\end{equation}
This choice ensures that there are no contributions from the 
symmetry breaking term $S$
to the energy-momentum tensor, such that $T_{ab}$ is entirely
determined by the supercurrent $V_a$.

The translation 
Ward operator and energy-momentum tensor are thus defined by
\begin{align}
\int \dx t^a(x) \,w_a^{(P)} &= \tfr{1}{2}\int\dV \Omega_P^\al(z) w_\al(A) +
c.c. \,, \\
\int \dx t^a(x) \partial^b T_{ab}(x) &=  \tfr{1}{32} \int\dV
\Omega_P^\al(z)  \bar D^\da V_{\al\da}(z) + c.c. \,. \label{Tabdef}
\end{align}
Integrating the trace identity (\ref{susych}) and its conjugate with $\Omega_P^\al$ and
differentiating with respect to $t^a(x)$, we obtain
the translation Ward identity
\begin{equation}
w_a^{(P)} \Gamma = -\partial^b T_{ab}
\end{equation}
(\ref{Tabdef}) relates the energy-momentum tensor $T_{ab}$ to the
$\theta\bar\theta $ component of $V_a$. With
\begin{equation}
V_a(z) = C_a(x) + \theta^\al \chi_{a\al}(x) + \bar\theta_\da
\bar\chi_a{}^\da + \theta\sigma^b \bar\theta \, v_{ab}(x) + \dots\, 
\label{Vcomp}
\end{equation}
we have
\begin{align}
T_{ab} &= \tfr{1}{8} (v_{ab} + v_{ba} - 2 \eta_{ab} v^c{}_c)\,, 
\label{Tabexplicit}\\
v_{ab} &= 4 (T_{ab} -\tfr{1}{3} \eta_{ab} T^c{}_c) +
\mbox{antisymmetric part}\,. 
\end{align}
Similarly  we obtain the conserved current $Q_{a \al}$ corresponding to rigid
supersymmetry by choosing a local transformation with vanishing Weyl
parameter $\sigma$ 
which reduces to rigid supersymmetry for constant parameters,
\begin{equation}
\Omega_Q^\al[q^\be(x), \bar q_\db(x) ] = - \bar\theta^2 q^\al(x) + 2
\theta^\al \bar \theta_\da \bar q^\da(x) + \tfr{3}{2} i \theta^2
\bar\theta^2 \sigma^{a\al\da} \pr_a \bar q_\da\,\,.
\end{equation}
Performing the same steps as in the calculation of $T_{ab}$ above, we
obtain the supersymmetry current in terms of the component fields of $V_a$,
\begin{align}
Q_a{}^\al &= i \left( \chi_a{}^\al - \sigma_{a\al\da} \sigma_b^{\be\da}
  \chi^b{}_\be \right)\,,\\
\chi_{a\al} &= -i \left( Q_{a\al} -\tfr{1}{3} Q^{b\be} \sigma_{b\be\da}
  \sigma_{a\al}{}^\da \right)\,. 
\end{align}
Finally, we obtain the $R$ current from 
\begin{equation}
\Omega_R^\al [r(x)] = -i \bar\theta^2 \theta^\al r(x) \, .
\label{OmegaR}
\end{equation}
However since for the $R$ transformation, the Weyl parameter $\sigma$
is non-zero even for rigid transformations,
it is not possible to have $\sigma[r(x)]=0$ here in contrast to
the two cases above.  
(\ref{OmegaR}) gives $\sigma[r(x)] = \tfr{2}{3}i r(x)$, where in the rigid case
$r$ is a constant. This yields the $R$ current
\begin{equation}
R_a(x) = C_a(x)\,,
\end{equation}
where $C_a(x)$ is the lowest $\theta$-component of $V_a$ in (\ref{Vcomp}).

\subsection{Component currents and corresponding background
 fields \label{vierbein}}

In the non-supersymmetric case of section 2, 
the energy-momentum tensor is given by
the derivative of the action with respect to the metric field
$g_{mn}$. In this section we show that this coupling  is also present
in the supersymmetric theory.
This may be seen in the component decomposition of the
prepotentials $H$ and $J$ and of the supercurrent $V$.
The energy-momentum tensor couples to components of both $H$ and $J$.
The corresponding two terms combine in just the right way to give the usual
coupling of the energy-momentum tensor to the metric.

On a non-supersymmetric manifold we may define the linearised vierbein
and its determinant by 
\begin{align}
e_a{}^m &= \delta_a{}^m + h_a{}^m + o(h^2)\,,\nonumber\\
e^{-1} &= (\det e_a{}^m)^{-1} = {\rm e}^{-{\rm tr} \log e_a{}^m} =
1-h^a{}_a + o(h^2)\,.\nonumber
\end{align}
In order to compare our formulation of curved superspace with the
non-supersymmetric case, we choose Wess-Zumino gauge such that the
prepotentials have the component structure
\begin{align}
H^b &= \theta \sigma^a \bar \theta h_a{}^b -i \bar \theta^2 \theta^\al
\Psi^b{}_\al + i \theta^2 \bar \theta_\da \bar \Psi^{b\da} + \theta^2 \bar
\theta^2 A^b \, , \\
\phi^3 &= \e^{-i\theta\sigma\bar\theta\partial} \left\{ e^{-1} \left(
    1-2i\theta^\al \sigma_{\al\da} \bar \Psi^{\al\da} + \theta^2 B \right)
\right\} \, , \\ 
J &=  \e^{-i\theta\sigma\bar\theta\partial} \left\{ -\tfr{1}{3} h^a{}_a
  -\tfr{2i}{3} \theta \sigma^a \bar \Psi_a + \tfr{1}{3} \theta^2 B \right\}\, .
\end{align}

In this formalism, the linearised vierbein $h_{ab}$ and the gravitino field
$\Psi^{a\al}$ appear in $H^{\al\da}$ as well as in $J$, $\bar J$.
In order to find the coupling of $h_{ab}$ to the
energy-momentum tensor and of $\Psi^{a\al}$ to the supersymmetry current,
we calculate the component decomposition of the effective action to first
order in $H$ and $J$,
\begin{equation}
\Geff^{(1)} = \tfr{1}{8} \int\dV H^{\al\da} V_{\al\da} + \int\dS J
\fdq{\Geff}{J}  + \int \dSb \bar J
\fdq{\Geff}{\bar J} \,.
\end{equation}
For the term involving $H$ we find to first order
\begin{align}
\tfr{1}{8} \int\dV H^{\al\da} V_{\al\da} = \int \dx \Bigl \{ & 
- h^{ab} \left( T_{ab} -\tfr{1}{3} h^a{}_a T^c{}_c + \tfr{1}{4}
  \epsilon_{abcd} \pr^cR^d \right) \nonumber \\
&+ \half \Psi^{a\al} \left( Q_{a\al} - \tfr{1}{3} Q^{b\ga}\sigma_{b\ga\da}
\sigma_{a\al}{}^\da \right) \nonumber \\
&+\half \bar \Psi^a{}_\da \left( \bar Q_a{}^\da -\tfr{1}{3} \bar Q^{b\dg}
  \sigma_{b\al\dg} \sigma_a^{\al\da} \right) \nonumber \\
&+ A^a R_a \,\, \Bigr\}
\label{HVcomp}
\end{align}
In the following we restrict ourselves to the on-shell case, $\delta \Gamma
/ \delta A = 0$. 
Using the definition of $w^\is$ (\ref{wis}), the Ward identity 
(\ref{swi}) and the trace equation (\ref{susych}),
the linearised term involving $J$ may be rewritten
as
\begin{align}
\int \dS J \fdq{\Geff}{J}  &= \int\dx \Bigl\{ D^2 J
  \fdq{\Geff}{J} + \tfr{3}{2} D^\al J \bar D^\da
  V_{\al\da} + \tfr{3}{4} J D^\al \bar D^\da V_{\al\da} \Bigr\} \nonumber
  \\
\intertext{This yields}
\int\dS  J \fdq{\Geff}{J}  + c.c. &= \int \dx \Bigl\{
-\tfr{1}{3} h^a{}_a T^c{}_c + \tfr{1}{6} \bar \Psi^a{}_\da \bar Q^{b\dg}
  \sigma_{b\al\dg} \sigma_a^{\al\da} + \tfr{1}{6} \Psi^{a\al} Q^{b\ga}
  \sigma_{b\ga\da} \sigma_{a\al}{}^\da\nonumber \\
& \qquad \qquad  - \tfr{4}{3} B \left. \fdq{\Geff}{J}
  \right|_0 -\tfr{4}{3} \bar B \left. \fdq{\Geff}{\bar J} \right|_0 \Bigr\}
\label{JScomp}
\end{align}
where $|_0$ denotes the zeroth $\theta$ component. 
Adding (\ref{HVcomp}) and (\ref{JScomp}) we obtain
\begin{align}
\Geff^{(1)} &= \int \dx \Bigl\{ -h^{(ab)} T_{ab} + \half \Psi^{a\al}
Q_{a\al} + \half \bar \Psi_{a\da} \bar Q^{a\da} + A^a R_a \nonumber \\
& \qquad \qquad \quad + 2 B \, S|_0 + 2 \bar B \, \bar S|_0 - \quar
h^{[ab]} \epsilon_{abcd}\pr^c R^d \Bigr\}  \,.
\end{align}

The metric tensor is given in terms of the inverse vierbein by
\[ g_{mn} = \eta_{ab} e_m{}^a e_n{}^b = \eta_{mn} - 2 h_{(mn)} +
o(h^2)\,,\]
such that the currents may be expressed as
\begin{gather}
T_{mn} = 2 \left. \fdq{\Geff}{g^{mn}} \right|_{\mbox{flat space}}  \, , \quad
Q_{a\al} = 2 \left. \fdq{\Geff}{\Psi^{a\al}} \right|_{\mbox{flat space}} 
 \, , \quad
R_a =  \left. \fdq{\Geff}{A^a} \right|_{\mbox{flat space}}  
\end{gather}
as expected. We note that while
the chiral compensator yields the trace of the energy-momentum tensor, the term
multiplying $h^{ab}$ on the r.h.s.~of (\ref{HVcomp}) 
is not trace-free due to the factor
of ${\ts \frac{1}{3}}$.  This is in contrast to the result of section
\ref{nonsusy} where the r.h.s.~of (\ref{nsc})
separates into a trace-free current and a trace term.

\subsection{Improvement terms}

The energy-momentum tensor of a real scalar field $\varphi$ on Minkowski
space, 
\begin{align}
T_{ab} &=  \left( \partial_a\varphi\partial_b \varphi
  - \half \eta_{ab}
  \partial^c\varphi \partial_c\varphi \right) +c  (\partial_a\partial_b -
\eta_{ab}\Box)\varphi^2 \,,
\end{align}
contains an improvement term $c  (\partial_a\partial_b -
\eta_{ab}\Box)\varphi^2$ whose divergence vanishes. The value $c=-\tfr{1}{6}$
is special in the 
sense that only for this value the energy-momentum tensor transforms
covariantly under conformal transformations. 

In the superfield approach, the energy-momentum tensor discussed
in section \ref{vierbein} is already improved in the sense that
the improvement term with $c=-\tfr{1}{6}$ is already contained in
(\ref{Tabexplicit}). This may be seen by expanding the dynamical field
$A$ in components,
\begin{equation}
A(z) = {\rm e}^{-i\theta\sigma\bar\theta\partial} (\varphi(x) + \theta\psi(x)
+ \theta^2 F(x))\,.\end{equation}
$T_{ab}$ is a sum of contributions from each of the component fields,
\begin{equation}
T_{ab} = T_{ab}^{(\varphi_1)} + T_{ab}^{(\varphi_2)} + T_{ab}^{(\psi)} +
T_{ab}^{(F)} \,,\end{equation}
where $\varphi_1$ and $\varphi_2$ are the real and imaginary part of the
complex scalar field $\varphi$.
For the massless Wess-Zumino model on curved superspace given by (\ref{Geff}),
we find from (\ref{Tabexplicit}) when restricting to flat space 
\begin{align}
T_{ab}^{(\varphi_1)} &=  z \left( \partial_a\varphi_1\partial_b \varphi_1
  - \half \eta_{ab}
  \partial^c\varphi_1 \partial_c\varphi_1 \right) 
- (\tfr{1}{6} z +
\tfr{2}{3}\xi) (\partial_a\partial_b -
\eta_{ab}\Box)\varphi_1^2 \,,\nonumber\\[0.2cm]
T_{ab}^{(\varphi_2)} &=  z \left( \partial_a\varphi_2\partial_b\varphi_2
  - \half \eta_{ab}
  \partial^c\varphi_2 \partial_c\varphi_2 \right) 
- (\tfr{1}{6} z -
\tfr{2}{3}\xi) (\partial_a\partial_b -
\eta_{ab}\Box)\varphi_2^2 \,. \label{Tphi}
\end{align}
In the classical approximation where $\xi=0$ for the $R$ invariant theory and
$z=1$, (\ref{Tphi}) yields
$c=-\tfr{1}{6}$. 
To higher orders, the coefficient $c$ acquires
renormalisation corrections. This behaviour is known from real
$\varphi^4$ theory \cite{KS1}. It is interesting however that in our case
the energy-momentum
tensor contributions for  the two real scalar
fields $\varphi_1$ and $\varphi_2$ have different renormalised improvement
coefficients.

\section{Discussion and Conclusion}

As far as renormalisation is concerned, the key feature of supersymmetric
theories on curved superspace is that there are two supergravity
prepotentials, of which $H_{\al \da}$, which couples to the supercurrent,
is Weyl invariant. Moreover the mass term, which for massless theories
appears as an auxiliary term within the 
Zimmermann approach to renormalisation, 
is independent of $H_{\al \da}$, such that 
no additional anomalies arise for two or even more supercurrent insertions.
This is in contrast to the non-supersymmetric case which has been studied in
\cite{KS1}, \cite{KS2}.
Thus for the massless Wess-Zumino model it is straightforward
to obtain Ward identities expressing the superconformal transformation 
properties of Green functions with an arbitrary number of insertions. 
For the renormalisation of $n$ supercurrent insertions, the effective
action requires counterterms in $H_{\al \da}$ up
to $n$th order. In our approach we have considered all orders in 
$H_{\al \da}$ in general, the counterterms being fixed by the
requirement of diffeomorphism invariance. The only remaining
freedom for the choice of counterterms amounts then to a
redefinition of $H_{\al \da}$ as a function of itself, which corresponds
to a shift of quasi-local terms between the Green functions and
the Ward identity.

For the case of the superconformal
transformation properties of the
supercurrent two point function we have determined
the contributions of the purely geometrical superconformal anomalies
explicitly. Here the situation is in close analogy to the 
energy-momentum tensor two point function in the non-supersymmetric case
in the sense that the Gau\ss-Bonnet
topological density does not contribute
to the superconformal transformation. 
At the fixed point,
the transformation of the two point function
is uniquely determined by the Weyl density in both cases.  
All geometrical
anomalies contribute to the supersymmetric trace of the
three point function $\bar D^\da_z \l V_{\al \da}(z) V_{\be \db}(z_1)
V_{\gamma \dg} (z_2) \r$, which is present in the Ward identity
obtained without imposing the superconformal constraint
$\bar D^\da \Omega^\al = D^\al \bar \Omega^\da$ on diffeomorphisms
and Weyl transformations.

The approach to double insertions presented here yields scheme-independent
results for the quasi-local terms in the supercurrent two point function
in a straightforward way. These terms appear to be of relevance 
to discussions of a possible extension of the Zamolodchikov C theorem
to higher dimensions \cite{Latorre}. We hope that our results presented
here, which are valid to all orders of perturbation theory,
may be helpful for a general treatment of such terms.  It has to
be kept in mind, though, that
here we have made use of some elegant features of supersymmetry as
far as renormalisation is concerned.  

\vspace{1cm}

\noindent {\bf Acknowledgement}\\
We are grateful to Prof.~Klaus Sibold for many useful discussions.

\appendix

\newpage

\renewcommand{\theequation}{\thesubsection.\arabic{equation}}

\section{Appendix}

\subsection{Local Superconformal Ward Operators}
\label{wardoperators}
\setcounter{equation}{0}
{\bf Diffeomorphisms ($\Lambda$ Transformations)}
\begin{align}
w_\alpha\il (A)&= \quar D_\alpha A \fdq{}{A}, \\[1ex]
w_\alpha\il (J)&= \quar D_\alpha J \fdq{}{J} 
- {\textstyle \frac{1}{12}} D_\al \fdq{}{J},\\[1ex]
w_\alpha\il (H)&= {}w_\alpha^{(0)}(H)\,+\,w_\alpha^{(1)}(H)\,+\,w_\alpha^{(2)}(H)\,+\,O(H^3),\\[1ex]
w_\alpha^{(0)}(H) &= \half \bar D^\da \fdq{}{H^{\alpha\da}}, \\[1ex]
w_\alpha^{(1)}(H) &= \quar \bar D^\da \left( \{ D_\alpha,\bar D_\da \}
         H^{\beta\db} \fdq{}{H^{\beta\db}} \right)
         +\quar \{D_\beta,\bar D_\db \} \bar D^\da \left( H^{\beta\db}
         \fdq{}{H^{\alpha\da}}\right) \nonumber\\
    &\quad  +\quar \bar D^2 \left( D_\alpha H^{\beta\db} \fdq{}{H^{\beta\db}} 
         \right),\\[1ex]
w_\al^{(2)}(H) &= {\textstyle \frac{1}{8}} \bar D^2 \{D_\gamma,\bar D_\dg\}
\left(H^{\gamma\dg} D_\al H^{\beta\db} \fdq{}{H^{\beta\db}}\right) \nonumber\\
&\quad +{\textstyle\frac{1}{24}} \bar D^\da \{D_\gamma,\bar D_\dg\}
 \left( H^{\gamma\dg} \{D_\beta,\bar D_\db\} \left( H^{\beta\db}\fdq{}{H^{\al\da}}\right)\right) \nonumber\\
&\quad +{\textstyle\frac{1}{12}} \bar D^\da \left(\{D_\al,\bar D_\da\} H^{\beta\db}
\{D_\gamma,\bar D_\dg\} \left(H^{\gamma\dg} \fdq{}{H^{\beta\db}}\right)\right)
\nonumber\\
& \quad + {\textstyle\frac{1}{24}} \bar D^\da \left( \{D_\al,\bar D_\da\} 
\left( H^{\gamma\dg} \{D_\gamma,\bar D_\dg\} H^{\beta\db} \right) \fdq{}{H^{\beta\db}}\right).
\end{align}

{\bf Super Weyl Transformations}
\begin{equation}
\begin{array}{rclcrcl}
w\is(A) &=& -A\fdq{}{A}, & \qquad & w_\al\is(A) &=&  -\frac{1}{12} D_\alpha \left(
A \fdq{}{A} \right), \\[1ex]
w\is(J) &=& \fdq{}{J}, & & w_\al\is(J) &= &\frac{1}{12} D_\al\fdq{}{J}, \\[1ex]
w\is(H)&=&0 \, , & & w_\al\is(H)&=&0 \, .
\end{array}
\end{equation}

{\bf Combined Transformations}
\begin{align}
w_\alpha (A)&= \quar D_\alpha A \fdq{}{A} -{\textstyle\frac{1}{12}} D_\alpha \left(
A \fdq{}{A} \right), \\[1ex]
w_\alpha (J)&= \quar D_\alpha J \fdq{}{J},\\[1ex]
w_\alpha (H)&= w_\al\il(H).
\end{align}

\subsection{Curvature terms}
\label{curvature}
\setcounter{equation}{0}
Here we give the expressions of the curvature tensors $W_{\al\be\ga}$,
$G_{\al\da}$, $R$, $\bar R$ to first order in the prepotentials $H_{\al\da}$,
$J$ and $\bar J$.
\begin{align}
R &= \bar D^2 \bar J + \tfr{1}{3} \bar D^2 \{ D_\al, \bar D_\da \} H^{\al\da}
 + \dots \\[0.2cm] 
\bar R &= D^2 J - \tfr{1}{3} D^2 \{ D_\al, \bar D_\da \} H^{\al\da}  +
 \dots\\[0.2cm]  
G_{\al\da} &= \{ D_\al, \bar D_\da\} (\bar J - J) + \half D \bar D^2 D
H_{\al\da} \nonumber \\
& \quad - \tfr{1}{6} [ D_\al, \bar D_\da] [D^\be, \bar D^\db] H_{\be\db} +
\half \{ D_\al, \bar D_\da \} \{ D^\be, \bar D^\db \} H_{\be\db}  +
 \dots\\[0.2cm]  
W_{\al\be\ga} &= \bar D^2 \{D_{(\al}, \bar D^\dg \} D_\be H_{\ga) \dg}  + \dots\\[0.2cm] 
\bar W_{\da\db\dg} &= D^2 \{D_\ga, \bar D_{(\da} \}\bar D_\db H^\ga{}_{\dg)} + \dots\,.\\
\intertext{These satisfy the Bianchi identities}
\Db^\da G_{\al\da} &= \half \D_\al R \\[0.2cm]
\D^\ga W_{\al\be\ga} &= \{\D_\al, \bar \D^\da \} G_{\be\da} + \{\D_\be, \bar
 \D^\da \} G_{\al\da}\,.
\end{align}

\subsection{Auxiliary expressions}
\setcounter{equation}{0}

\label{appdoubleins}
\begin{align}
\Delta^{(2)}_{\al\be\db\ga\dg}{}^{\de\dd} (z;z_1,z_2) \bigl( \,\bullet\, \bigr)
&=
- \delta_\ga^\de \delta_\dg^\dd \bar D^2 \{ D_\be, \bar D_\db \} \left(
  \delta^8(z-z_1) D_\al \delta^8(z-z_2) \,\bullet\, \right) \nonumber \\
& \quad - \tfr{1}{3} \delta_\al^\de \bar D^\dd \{D_\be, \bar D_\db \}
\left( \delta^8(z-z_1) \{ D_\ga, \bar D_\dg \} (\delta^8(z-z_2) \,\bullet\, )
\right) \nonumber \\ 
& \quad -\tfr{2}{3} \delta_\be^\de \delta_\db^\dd \bar D^\da \left(
  \{D_\al, \bar D_\da \} \delta^8(z-z_1) \{ D_\ga, \bar D_\dg \}
  (\delta^8(z-z_2) \,\bullet\, ) \right) \nonumber \\
& \quad -\tfr{1}{3} \delta_\ga^\de \delta_\dg^\dd \bar D^\da \left( \{
  D_\al, \bar D_\da \} (\delta^8(z-z_1) \{ D_\be, \bar D_\db \} \delta^8
  (z-z_2)) \,\bullet\, \right) \nonumber \\
& \quad + \left( (\be\db,z_1) \leftrightarrow (\ga\dg,z_2) \right) \,,
\end{align}

\begin{align}
L_{\al\da\be\db}(z-z_1, z-z_2) &=  -\tfr{4}{3} \delta^8(z-z_1) \pr_{\al\da}
\pr_{\be\db} \bar D^2 \delta^8(z-z_2) \nonumber\\
&\quad  -\tfr{4}{9} \pr_{\al\da} ( \delta^8(z-z_1) \pr_{\be\db} \bar
D^2 \delta^8(z-z_2)) \nonumber \\
& \quad + \left((\al\da, z_1) \leftrightarrow (\be\db, z_2)\right) \,,
\\[1.5ex]
M^a{}_{\epsilon,\al\da\be\db}(z-z_1,z-z_2) &=
-2i \sigma^{a\phi}{}_\da \delta^8(z-z_1) \delta W_{\al\phi\epsilon,\be\db}(z-z_2)
\nonumber \\& \quad 
-4i \sigma^a_{\epsilon\dg} \delta^8(z-z_1) \bar D^\dg \delta
G_{\al\da\be\db}(z-z_2)  \nonumber \\[0.5ex]
& \quad +\tfr{4}{3} i \sigma^a_{\ga\dg} \varepsilon_{\alpha\epsilon}
\delta^8(z-z_1) \bar D_\da \delta G^{\ga\dg}{}_{\be\db}(z-z_2) 
\nonumber\\& \quad 
-\tfr{4}{9} \sigma^a_{\ga\da} \varepsilon_{\al\epsilon}
\delta^8(z-z_1) D^\ga \bar D^2 \pr_{\be\db} \delta^8(z-z_2) \nonumber\\[0.5ex]
& \quad +\tfr{8}{9} \sigma^a_{\al\da} \delta^8(z-z_1) D_\epsilon \bar D^2
\pr_{\be\db} \delta^8(z-z_2) 
\nonumber \\& \quad 
-\tfr{4}{9} \sigma^a_{\epsilon \da} \delta^8(z-z_1) D_\al \bar D^2
\pr_{\be\db} \delta^8(z-z_1) \nonumber \\[0.5ex]
& \quad + \left( (\al\da, z_1) \leftrightarrow (\be\db,z_2) \right) \,,
\\[1.5ex]
N^{\ga\dg}{}_{\al\da}(z-z_1, z-z_2) &=  \tfr{8}{3} \delta^8(z-z_1) \delta
G^{\ga\dg}{}_{\be\db}(z-z_1) \,,\\
\intertext{where}
\delta W_{\al\phi\epsilon,\be\db} (z-z') &\equiv \left. \fdq{}{H^{\be\db}(z')}
W_{\al\phi\epsilon} \right|_{H=J=0}  \,.
\end{align}

\end{document}